\documentclass[a4paper,12pt]{article}

\usepackage{geometry,slashed}
\usepackage{amssymb}

\RequirePackage[T1]{fontenc} 

\RequirePackage{graphicx}
\RequirePackage{flushend}
\RequirePackage[numbers,sort&compress]{natbib}
\RequirePackage[colorlinks,citecolor=blue,urlcolor=blue,linkcolor=blue]{hyperref}

\usepackage{yfonts}
\usepackage{booktabs}
  \usepackage{soul}
  \usepackage[usenames,dvipsnames]{xcolor}
 
 \definecolor{X575}{rgb}{0.05, 0.7, 0.05}
 
\usepackage{cancel} 
\usepackage{graphicx} 
 \usepackage{pgffor}
\usepackage{array}
\usepackage{amsmath}
 \usepackage{color}
 \usepackage{hyperref}
 \usepackage{authblk}
 \usepackage{epstopdf}
 \usepackage{enumerate}
 \usepackage{multirow}
 \usepackage{siunitx}
 \usepackage{float}

  \usepackage{listings}
  \usepackage{calrsfs}
  \usepackage{subfigure}

\graphicspath{{./figures/}}
  
  \DeclareMathAlphabet{\pazocal}{OMS}{zplm}{m}{n}

\newcommand{\gev}{\textrm{GeV}}

\newcommand{\Hbb}{H b \bar b}

\newcommand{\mb}{m_b}
\newcommand{\mt}{m_t}
\newcommand{\mz}{m_Z}
\newcommand{\mw}{m_W}
\newcommand{\mh}{m_H}
\newcommand{\mbpole}{m_{b}^{\rm pole}}
\newcommand{\mbMSbar}{m_{b}^{\MSbar}}
\newcommand{\mglong}{{\sc\small MadGraph5\_aMC@NLO}}
\newcommand{\mgshort}{{\sc\small MG5\_aMC}}

\newcommand{\gamt}{\Gamma_t}
\newcommand{\gamw}{\Gamma_W}
\newcommand{\gamz}{\Gamma_Z}
\newcommand{\gamh}{\Gamma_H}
\newcommand{\gmu}{G_\mu}

\newcommand\mf{{\sc\small MadFKS}}
\newcommand\ml{{\sc\small MadLoop}}
\newcommand\ct{{\sc\small CutTools}}
\newcommand\nin{{\sc\small Ninja}}
\newcommand\coll{{\sc\small Collier}}

\chardef\MyArticleWithColor=\pdfcolorstackinit page direct{0 g}

\def\alphas{\alpha_s}

\def\TO{{\rightarrow}}
\def\ord{\mathcal{O}}
\def\alphas{\alpha_s}
\def\LO{{\rm LO}}
\def\NLO{{\rm NLO}}
\def\NLOQCD{\rm NLO_{QCD}}
\def\LOQCD{\rm LO_{QCD}}

\def\NLOQCD{\rm NLO_{QCD}}
\def\NLOQCDEW{\rm NLO_{QCD+EW}}
\def\CNLO{\rm NLO_{all}}

\newcommand{\MSbar}{{\rm \overline{MS}}}

\begin{document}

\title{RIP $\Hbb$: \\
How other Higgs production modes conspire to kill a rare signal at the LHC}

\author[1]{Davide Pagani\thanks{davide.pagani@desy.de}}
\author[2]{Hua-Sheng Shao\thanks{huasheng.shao@lpthe.jussieu.fr}}
\author[3]{Marco Zaro\thanks{marco.zaro@mi.infn.it}}

\affil[1]{\small DESY, Theory Group, Notkestrasse 85, 22607 Hamburg, Germany}
\affil[2]{\small Laboratoire de Physique Th\'eorique et Hautes Energies (LPTHE), UMR 7589, Sorbonne Universit\'e et CNRS, 4 place Jussieu, 75252 Paris Cedex 05, France}
\affil[3]{\small INFN, Sezione di Milano \& TIFLab, Via Celoria 16, 20133 Milano, Italy}

\date{}

\maketitle

  \vspace*{-9cm}
  {
      {{\color{blue}{DESY 20-089}  
      \hfill \color{blue}{TIF-UNIMI-2020-16}}
  }
  \vspace*{9cm}

\begin{abstract}
The hadroproduction of a Higgs boson in association with a bottom-quark  pair ($H b \bar b$) is commonly considered as the key process for directly probing the Yukawa interaction between the Higgs boson and the bottom quark ($y_b$). However, in the Standard-Model (SM) this process is also known to suffer from very large irreducible 
backgrounds from other Higgs production channels, notably gluon-fusion ($gg$F). In this paper we calculate for the first time the so-called QCD and electroweak complete-NLO predictions for $H b \bar b$ production, using the four-flavour scheme. Our calculation shows that not only the $gg$F but also the $ZH$ and even the vector-boson-fusion channels are sizeable irreducible 
backgrounds. Moreover, we demonstrate that, at the LHC, the rates of these backgrounds are very large with respect to the ``genuine'' and $y_b$-dependent $H b \bar b$ production mode. In particular, no suppression occurs at the differential level and therefore backgrounds survive typical analysis cuts. This fact further jeopardises
the chances of measuring at the LHC the $y_b$-dependent component of $\Hbb$ production in the SM. Especially, unless  $y_b$  is significantly enlarged by new physics, even for beyond-the SM scenarios the direct determination of $y_b$ via this process seems to be hopeless at the LHC.

\end{abstract}

\section{Introduction}
\label{sec:intro}

Since the beginning of its operation the Large Hadron Collider (LHC) has disclosed an unprecedented amount of information on the nature of the Higgs boson.  After having discovered this particle \cite{Aad:2012tfa,Chatrchyan:2012xdj}, the ATLAS and CMS collaborations have also observed all the four main production modes \cite{Aad:2012tfa,Chatrchyan:2012xdj, Sirunyan:2017khh, ATLAS:2018lur, Sirunyan:2018kst, Aaboud:2018zhk, Aaboud:2018urx, Sirunyan:2018hoz}: gluon-fusion ($gg$F), vector-boson-fusion (VBF), vector-boson associated  production ($VH$), and top-quark pair associated production ($t \bar t H$). Moreover, many of the decay channels, $\gamma\gamma, VV^*, \tau^+\tau^-$ and $b\bar b$  have been observed, too. All these measurements has lead to a very clear statement: the properties of this particle are compatible with those of the Higgs boson of the Standard Model (SM) of elementary particles. Notably, the couplings of the Higgs boson have already been found to be compatible with the SM predictions at 15-20\% accuracy for the case of the third-generation fermions and even at $\sim5\%$ accuracy for the $W$ and $Z$ bosons~\cite{Aad:2019mbh, ATLAS:2020qdt}.

On the other hand, there is still large room for possible beyond-the-SM (BSM) effects. For instance, with the exception of the muon~\cite{Aaboud:2017ojs, Sirunyan:2018hbu, CMS:2020eni, Aad:2020xfq}, couplings to the first two fermion generations are largely unconstrained. A similar situation is also present at the moment  for the Higgs self coupling, for which only bounds of the order of 5-10 times the SM value have been obtained~\cite{Aad:2019uzh, ATLAS:2019pbo}. In addition, the extracted value for a specific Higgs coupling  depends on the assumptions on (the structure of) other interactions, both in the Higgs sector as well as in the top-quark or electroweak (EW) sector. For instance, Higgs production processes at the LHC are always measured in conjunction with specific decay channels, including those still unconstrained. Therefore the extraction of the couplings in a given production mode is affected by those appearing in the decay channel, and vice versa. Moreover, branching ratios depend on the total decay width and in turn on all the other decay channels.
For this reason, having independent measurements of various production processes can certainly improve the  precision in the determination of a specific coupling, and meanwhile it allows to relax the underlying theoretical assumptions.

At the LHC, in the case of the bottom-quark Yukawa coupling ($y_b$), a direct sensitivity can be in principle achieved both via the $H\TO b \bar b$  decay or via the associate production of a Higgs boson together with a $b \bar b$ pair.
While the $H\TO b \bar b$ decay has already been observed  in conjunction with $VH$ production \cite{Aaboud:2018zhk, Sirunyan:2018kst, Aaboud:2019nan, ATLAS:2020udg, Aad:2020jym, Aad:2020eiv}, no dedicated SM analysis has ever been performed by 
the ATLAS and CMS collaboration in order to measure the $\Hbb$ production process. Indeed, this process has an inclusive cross section that is comparable to the one of $t \bar t H$ production ({\it e.g.} $\sim0.5$ pb at 13 TeV). However, at variance with $t \bar t H$ production, at least one $b$-jet has to be tagged in order to make it distinguishable from the inclusive Higgs-boson production process, whose production rate is $\sim100$ times larger than $\Hbb$ production alone. The problem is that tagging a $b$-jet dramatically reduces the cross section, but as said this is an unavoidable procedure for  obtaining a possible {\it direct} sensitivity on the bottom-quark Yukawa coupling. Without tagging any $b$-jet, even bottom-quark loops in $gg$F, which induce an {\it indirect} sensitivity on $y_b$,  have a larger contribution: about $-6$\% for the inclusive cross section and up to
 $-10$\% for the Higgs boson at small transverse momentum~\cite{Mantler:2012bj, Grazzini:2013mca,Banfi:2013eda,Bagnaschi:2011tu,Frederix:2016cnl,Bagnaschi:2015bop,Mantler:2015vba,Harlander:2014uea}. 
 
In the past, $\Hbb$ production has nevertheless received a lot of attention from the theoretical community. Indeed, in BSM theories with an extended Higgs sector, such as the  two-Higgs-doublet-model (2HDM) or the minimal-supersymmetric-SM (MSSM), the coupling of the Higgs boson to bottom quarks can be significantly enhanced. Moreover, in the context of higher-order QCD corrections, this process is particularly interesting also from a formal point of view. Featuring bottom quarks in the hard process,  two different schemes can be adopted when performing perturbative calculations: the so-called four-flavour scheme (4FS) and five-flavour scheme (5FS). In the former, the bottom quark is considered as massive, while in the latter the bottom mass is set equal to zero.

In the 4FS, next-to-leading order (NLO) QCD predictions were firstly obtained in Refs.~\cite{Dittmaier:2003ej,Dawson:2003kb}, then for MSSM-type couplings in Ref.~\cite{Dawson:2005vi}, and including supersymmetry (SUSY) QCD corrections in Ref.~\cite{Liu:2012qu,Dittmaier:2014sva}. In the 5FS, many more results are present in the literature, since the higher-order perturbative calculations are technically much easier, owing to the smaller number of final-state particles. Indeed, corrections up to next-to-next-to-next-to-leading order (N$^3$LO) QCD accuracy~\cite{Dicus:1998hs,Balazs:1998sb,Harlander:2003ai,Duhr:2019kwi} are available. Distributions at the parton level were obtained for $H$+$b$ and $H$+jet  production at NLO in Refs.~\cite{Campbell:2002zm,Harlander:2010cz}. Next-to-next-to-leading order (NNLO) accuracy has been then reached for jet rates in Ref.~\cite{Harlander:2011fx} and for fully differential distributions in Ref.~\cite{Buehler:2012cu}.
The  spectrum of the Higgs boson transverse-momentum  was studied analytically
up to $\mathcal{O}(\alpha_s^2)$ in Ref.~\cite{Ozeren:2010qp}, while
resumming at NLO+NLL and NNLO+NNLL  accuracy  in Refs.~\cite{Belyaev:2005bs} and \cite{Harlander:2014hya}, respectively.
For what concerns NLO QCD predictions matched to parton shower effects (NLO+PS),  both the 4FS and the 5FS cases were studied for the first time within {\sc MadGraph5\_aMC@NLO} \cite{Wiesemann:2014ioa} and subsequently also via the {\sc Powheg} \cite{Jager:2015hka} and {\sc Sherpa} \cite{Krauss:2016orf} approaches. 
Finally, at the level of the total cross section, differences between results obtained in the two schemes have been studied in Refs.~\cite{Maltoni:2012pa,Lim:2016wjo} and then
 the state-of-the-art 4FS and 5FS predictions have been combined in Refs.~\cite{Forte:2015hba,Forte:2016sja,Bonvini:2015pxa,Bonvini:2016fgf} and very recently in Ref.~\cite{Duhr:2020kzd}, which matches
 the 5FS prediction at N$^3$LO QCD accuracy and the 4FS prediction at NLO QCD.

The difficulties in the extraction of $y_b$ via the measurement of $\Hbb$ production do not originate only from its very small cross section. In the 4FS, NLO QCD corrections to $\Hbb$ interfere with gluon-fusion Higgs production (at LO) with an extra emission of a $b \bar b$ pair ($gg$F+$b\bar b$). This interference leads to a term proportional to $y_b y_t$, where $y_t$ is the top-quark Yukawa coupling \cite{Wiesemann:2014ioa}. Moreover, the  LO diagrams of $gg$F+$b\bar b$ are infrared (IR) finite and lead to terms proportional to $y_t^2$, when squared. The $y_t y_b$ term is non-negligible w.r.t.~the term proportional to  $y_b^2$ originating from the ``genuine'' $\Hbb$ production, and especially the $y_t^2$ term is much larger than the $y_b^2$ contribution.  Both the  $y_t y_b$ and $y_t^2$ terms have been calculated at NLO QCD accuracy in Ref.~\cite{Deutschmann:2018avk}; if at least one $b$-jet is required, the $y_t y_b$ term is $\sim - 20\%$ of the  $y_b^2$ term, while the $y_t^2$ term is $\sim 5$ times the  $y_b^2$ term. 
In this paper, we calculate the cross section of Higgs boson production in association with a $b \bar b$ pair, $\Hbb$, at ``complete-NLO'' accuracy in the 4FS. In other words, we compute all the SM contributions from QCD and EW origin at the tree and one-loop level. 
The choice of the 4FS is driven by the  fact that when EW corrections are taken into account, the SM relation between $y_b$ and $m_b$ cannot be ignored and therefore in the 5FS one must enforce $y_b=0$.
 Complete-NLO predictions take into account not only the LO (order  $\alphas^2\alpha$), the NLO QCD  (order  $\alphas^3\alpha$) and NLO EW  (order  $\alphas^2\alpha^2$) corrections, but also all the possible terms of order $\alpha_s^m \alpha^{n+1}$ with $m,n\ge 0$ and $m+n=2,3$. Part of the 
NLO EW corrections (only the gluon-gluon initial-state contribution) has already been calculated also in Ref.~\cite{Zhang:2017mdz}. By considering complete-NLO predictions, new topologies open up on top of the aforementioned $gg$F+$b\bar b$ one. Terms with $n\ge 2$ include $ZH$ production, with subsequent $Z$ decay into a $b \bar b$ pair. Moreover, at order  $\alphas\alpha^3$ even VBF configurations with $Z$ bosons arise.  In this paper we show, for the first time,  that the numerical impact of these additional contributions is sizeable and sometimes even dominate, though they are formally suppressed by the small EW coupling constant. Especially, we show that the suppression of their relative contributions via {\it ad hoc} cuts inevitably strongly reduces also the total rates.

In our study, we demonstrate  that the idea of directly extracting $y_b$ from the measurement of $\Hbb$ at the LHC is substantially hopeless. 
The rates for this process are small and contaminated by terms that depend on $y_t$ and the $HZZ$ coupling. Reducing this contamination implies also a strong reduction of the cross section of the term depending only on $y_b$.

The aforementioned computation has been performed via the latest version of {\sc\small Mad\-Graph5\_\-aMC@NLO}~\cite{Frederix:2018nkq}, which is public and has been extended in order to be able to calculate NLO EW corrections, and in general complete-NLO predictions, also in the 4FS.\footnote{This new capability will be included in a future release of {\mglong}}  Since the results presented in this paper
represent the first complete-NLO (and also NLO EW) computation performed in such a scheme, we will also discuss in the text the relevant technical aspects connected to the usage of the 
4FS in NLO EW corrections.

The paper is organised as follows. In Sec.~\ref{sec:setup} we describe our calculation set-up. The technical improvements performed for calculating  complete-NLO predictions in the 4FS via {\sc\small Mad\-Graph5\_\-aMC@NLO} are documented in Sec.~\ref{sec:model}. In Sec.~\ref{sec:topologies} we describe the different topologies entering our calculation at different perturbative orders and in Sec.~\ref{sec:msbar} we discuss the problems related to the $\MSbar$ renormalisation of $y_b$, when EW corrections are present. Numerical results are presented in Sec.~\ref{sec:results}. Input parameters are given in Sec.~\ref{sec:inputs}, while numerical results at the inclusive and differential level are presented and commented in details in Secs.~\ref{sec:inclusive} and \ref{sec:differential}, respectively. The main phenomenological result of our work, {\it i.e.} the fact that the idea of directly extracting $y_b$ from the measurement of $\Hbb$ at the LHC is substantially hopeless, is motivated in detail in Sec.~\ref{sec:inclusive} and further corroborated by the analysis at the differential level in Sec.~\ref{sec:differential}. We give our conclusions and outlook in Sec.~\ref{sec:conclusions}.

\section{Complete-NLO predictions for $\Hbb$ production }\label{sec:setup}

As mentioned in Sec.~\ref{sec:intro}, in this work we present the complete-NLO  predictions for $\Hbb$ production at the LHC, namely, we exactly take into account all the one-loop  and real-emission corrections of QCD and EW origin. Expanding in powers of $\alpha_s$ and $\alpha$, the first non-vanishing contribution to the cross section of $\Hbb$ production is of $\ord(\alpha_s^2 \alpha)$, and it is induced by  tree-level $gg\rightarrow \Hbb$ and $q \bar q \rightarrow \Hbb$ diagrams. Complete-NLO predictions for $\Hbb$ production include all the $\ord(\alpha_s^m \alpha^{n+1})$ contributions with $m,n\ge 0$ and $m+n=2,3$. Following the notation already used in Refs.~\cite{Frixione:2014qaa, Frixione:2015zaa, Pagani:2016caq, Frederix:2016ost, Czakon:2017wor, Frederix:2017wme, Frederix:2018nkq, Broggio:2019ewu, Frederix:2019ubd}, the different contributions to any differential or inclusive cross section $\Sigma$ can be denoted as:
\begin{align}
\Sigma^{}_{\LO}(\alpha_s,\alpha) &= \alpha_s^2 \alpha \Sigma_{3,0}^{} + \alpha_s \alpha^2 \Sigma_{3,1}^{} + \alpha^3 \Sigma_{3,2}^{} \nonumber\\
 &\equiv \LO_1 + \LO_2 + \LO_3\, , \label{eq:blobLO} \\
 \Sigma^{}_{\NLO}(\alpha_s,\alpha) &= \alpha_s^3 \alpha \Sigma_{4,0}^{} + \alpha_s^2 \alpha^2 \Sigma_{4,1}^{} + \alphas \alpha^3 \Sigma_{4,2}^{}+ \alpha^4 \Sigma_{4,3}^{} \nonumber\\
 &\equiv \NLO_1 + \NLO_2 + \NLO_3+  \NLO_4\, . \label{eq:blobNLO} 
\end{align}

The ``standard'' LO contribution is in our notation the $\LO_1$, while the term ``$\LO$'' is rather used for denoting the sum of all  $\LO_i$ terms, consistently with Eq.~\eqref{eq:blobLO}. Similarly, according to Eq.~\eqref{eq:blobNLO}, the term $\NLO$ is used to denote the sum of all  $\NLO_i$ terms. Thus, the quantity $\LO+\NLO$ corresponds to the complete-NLO predictions. Further definitions will be given also later in the text in Eqs.~\eqref{eq:def1}--\eqref{eq:def5}; they constitute the quantities entering the phenomenological discussion of Sec.~\ref{sec:results}.

The calculation has been performed via the latest version of the public code {\sc\small Mad\-Graph5\-\_aMC@NLO}~\cite{Frederix:2018nkq} (\mgshort\ hence-force), which enables the user to 
compute predictions at NLO EW and complete-NLO accuracy also in the 4FS, for a generic process in the SM or in any model for which
the necessary counterterms are known. No {\it ad hoc} customisation of the code has been put in place in order to perform this specific calculation for the $\Hbb$ final state.

 Technical details about the evaluation of loops and the ultraviolet (UV) renormalisation procedure for EW corrections in the 4FS are explained in Sec.~\ref{sec:model}. The IR singularities, in the  \mgshort\ framework, \cite{Alwall:2014hca}, are dealt with via the FKS method~\cite{Frixione:1995ms, Frixione:1997np}, which  
 is automated in the module \mf~\cite{Frederix:2009yq,
Frederix:2016rdc}. From the FKS side, the usage of the 4FS in conjunction with EW corrections does not pose additional difficulties; bottom quarks are massive and
do not give rise to any collinear singularity. In practice, they are treated in the same way as top quarks. Initial-state bottom quarks are not present and IR divergences can be regulated according to the procedure explained in Sec.~3 of Ref.~\cite{Frederix:2018nkq}.

\subsection{Complex mass scheme renormalisation and virtual matrix-element evaluation in the 4FS}\label{sec:model}

In order to handle the intermediate resonances appearing in the Feynman diagrams, we  adopt in our calculation the well-known complex mass scheme~\cite{Denner:1999gp,Denner:2005fg}, in which one has to modify the renormalisation conditions yielding complex-valued renormalised parameters. Beyond LO, the carry out of the complex renormalisation procedure becomes subtle and requires very careful treatments. In particular, Ref.~\cite{Frederix:2018nkq} has explored several important issues related to the computations at NLO in general.
However, before this present publication, in the \mgshort\ framework NLO EW corrections and more in general complete-NLO predictions could be performed only in the 5FS, {\it i.e.}, with massless bottom quarks. In order to perform the calculation of the complete-NLO predictions of any SM process in the 4FS, in particular $\Hbb$ as a case study in the present paper, we have extended the capabilities of \mgshort. Therefore, in this work we report also the new feature of \mgshort: NLO EW corrections and more in general complete-NLO predictions can also be obtained in the 4FS. 

Technical difficulties, as already pointed out in Ref.~\cite{Frederix:2018nkq}, are related to the presence of very different scales in the process and the use of the complex renormalisation conditions in it.
 The main concern here is that, since the complication arises from multiple-scale Feynman integrals, we have to take care of the analytic continuation from the first Riemann sheet to other sheets of two-point one-loop integrals appearing in the mass and wave-function UV renormalisation counterterms in the complex mass scheme. Instead of performing Taylor expansions in the simplified version ({\it cf.}, {\it e.g.}, Sec.~6.6.3 in Ref.~\cite{Denner:2019vbn}), our implementation follows a more rigorous approach, the so-called {\it trajectory approach}, first proposed in Ref.~\cite{Frederix:2018nkq}. The latter does not introduce additional approximations, following the original spirit of the complex mass scheme. However, the difference between the two is in general formally beyond NLO in the SM and thus numerically insignificant~\cite{Bendavid:2018nar} for the SM particle's mass spectrum. On the contrast, for a general mass spectrum in, {\it e.g.}, a BSM theory, the simplified version could fail to produce the correct result, while the trajectory approach always selects the correct Riemann sheets for the multivalued complex functions. The concrete realisation of the trajectory approach in our implementation has been given in the Appendix E.2 of Ref.~\cite{Frederix:2018nkq}. More specifically, the numerical routines follow Eqs.~(E.44)--(E.47) in that appendix. We want to stress that although the general conceptual issues have already been extensively discussed in Ref.~\cite{Frederix:2018nkq}, the novel aspect in the current paper is the first complex mass scheme realisation of 4FS in \mgshort, as well as its validation\footnote{Several internal validations have been done from different angles. In particular, a systematic test introduced in the Appendix E.1 of Ref.~\cite{Frederix:2018nkq} has been performed.} in the context of the general trajectory approach. Such an implementation will be publicly available in a future  release  of {\mgshort}. 

The evaluation of one-loop virtual matrix elements in {\mgshort} is performed in the \ml\ module~\cite{Hirschi:2011pa,
Alwall:2014hca}, by using
different types of techniques for Feynman one-loop integral reduction, namely, the integrand reduction ({\it e.g.}, the so-called OPP~\cite{Ossola:2006us} and
Laurent-series expansion~\cite{Mastrolia:2012bu} methods)
or tensor integral reduction~\cite{Passarino:1978jh,Davydychev:1991va,Denner:2005nn} approaches. 
\ml\ is used to automatically generate the one-loop renormalised amplitudes and to evaluate them via dynamically switching among the different one-loop integral techniques by employing the public codes \ct~\cite{Ossola:2007ax}, \nin~\cite{Peraro:2014cba,
Hirschi:2016mdz} and \coll~\cite{Denner:2016kdg}. An independent in-house implementation of  the {\sc OpenLoops} optimisation~\cite{Cascioli:2011va} enables the boosting of the fast evaluations of the virtual matrix elements.
Contrary to the 5FS, in the 4FS a worry may come from the possible numerical instability occurring in one-loop evaluations due to the smallness of the bottom quark. As an example, in our calculation ({\it i.e.} complete-NLO predictions for $\Hbb$ production) we find that the self-diagnostic and recovery strategies implemented in \ml5~\cite{Alwall:2014hca} are already quite effective. Namely, $99.79\%$ phase-space points have been successfully calculated by \nin\ in the double precision, and the majority of the remaining unstable phase-space points have been successfully rescued by \coll\ and \ct\ in the double precision, while only two phase-space points (amounting to one in hundred million events) have needed the quadruple precision architecture. No event has failed to be rescued.

\subsection{Topologies contributing to the $\Hbb$ final state}
\label{sec:topologies}

\begin{figure}[!t]
    \centering
    \subfigure[]
    {
        \includegraphics[width=4cm, height=3.5cm]{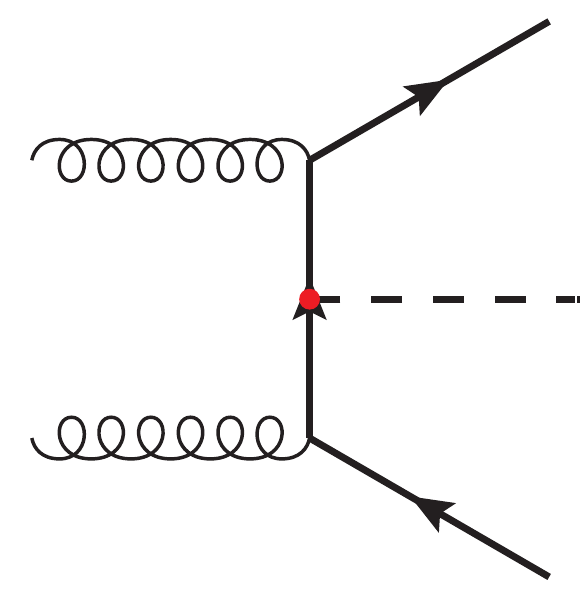}
        \label{fig:ggHbb}
    }
    \subfigure[]
    {
        \includegraphics[width=4cm, height=3.5cm]{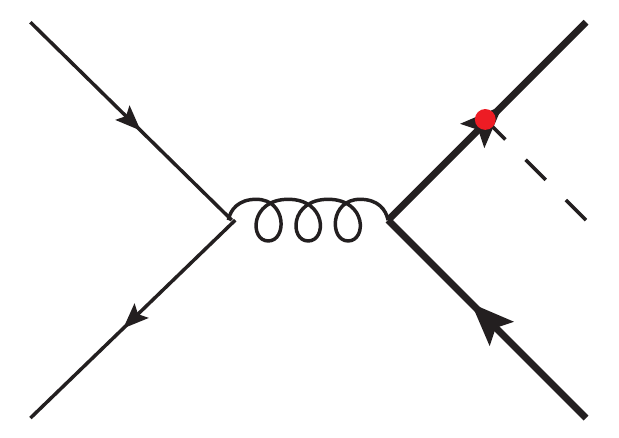}
        \label{fig:qqHbb}
    }
    \subfigure[]
    {
        \includegraphics[width=4cm, height=3.5cm]{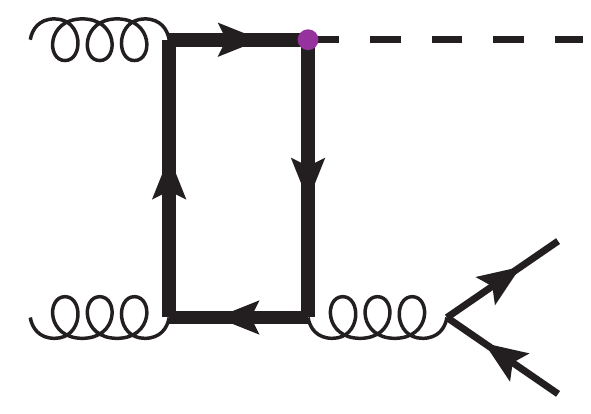}
        \label{fig:Yt}
    }
            \subfigure[]
    {
        \includegraphics[width=4cm, height=3.5cm]{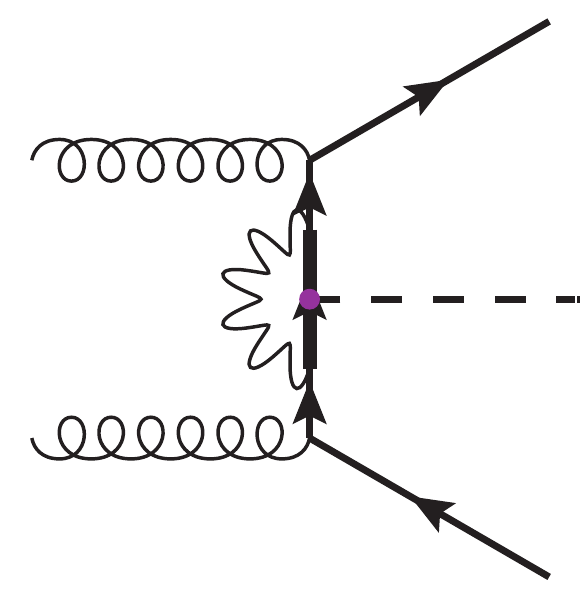}
        \label{fig:gg2hbb_yt}
    }
    \subfigure[]
    {
        \includegraphics[width=4cm, height=3.5cm]{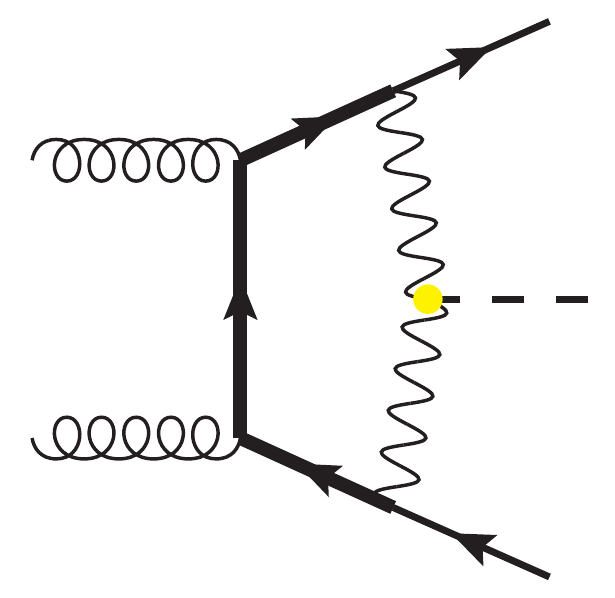}
        \label{fig:ggHbb_W}
    }
            \subfigure[]
    {
        \includegraphics[width=4cm, height=3.5cm]{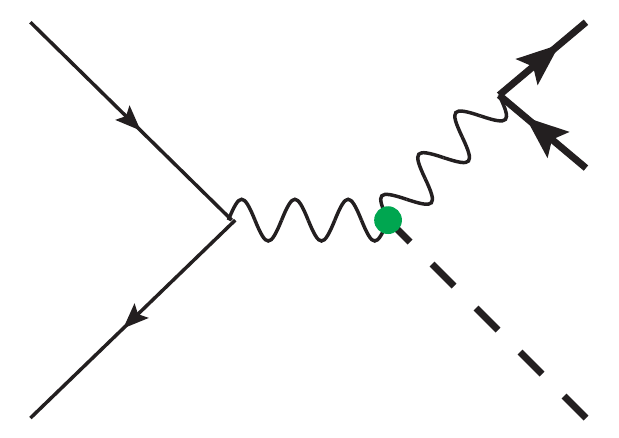}
        \label{fig:VH}
    }

    \subfigure[]
    {
        \includegraphics[width=4cm, height=3.5cm]{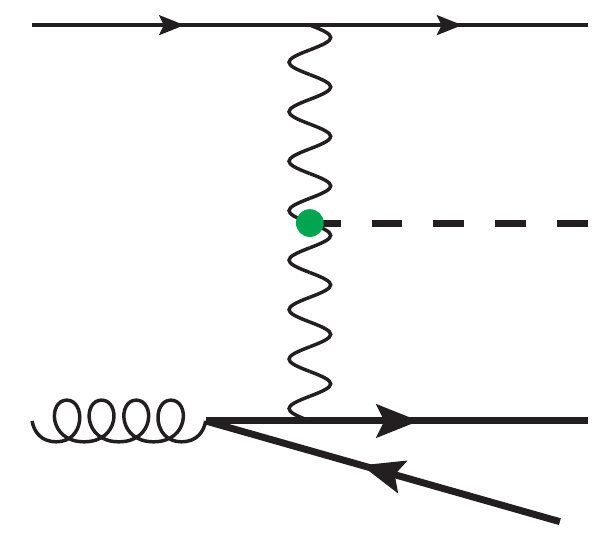}
        \label{fig:VBF}
    }
    \caption{Representative Feynman diagrams appearing in the complete-NLO calculation for $\Hbb$ production. The thick, medium-thick and thin solid lines represent the top, bottom and light (anti-)quarks, respectively. The dashed lines stand for the Higgs boson, the curly lines are gluons, and the wiggly lines are the weak bosons ($W$ and $Z$). The red/violet/green/yellow bullets represent $\Hbb$/$Ht\bar{t}$/$HZZ$/$HWW$ interactions.}
    \label{fig:sample_subfigures}
\end{figure}

In order to better understand the results of our calculation, it is first of all useful to describe the various topologies of the diagrams entering each perturbative order of the complete-NLO predictions, as summarised in Tab.~\ref{table:orders}.
The $\LO_1$ originates from the ``genuine'' $\Hbb$ production process in the 4FS, {\it i.e.}, topologies  that feature a bottom-Higgs coupling, such as the one depicted in Fig.~\ref{fig:ggHbb} for $gg\TO \Hbb$. Also contributions from the quark-antiquark initial state are present at this order, Fig.~\ref{fig:qqHbb},  but their contribution is much smaller than those form the gluon-gluon initial state. The $\LO_3$ receives contributions from  ``genuine'' $\Hbb$ production via $\gamma\gamma\TO \Hbb$ diagrams, but the $q \bar q \TO \Hbb$ diagrams dominate at this order. Indeed, not only the $\gamma\gamma\TO \Hbb$ process is suppressed by the photon PDF,  the $q \bar q \TO \Hbb$ diagrams contain an additional topology: $q \bar q \TO ZH$ production with the subsequent $Z \TO b \bar b$ decay, see the diagram illustrated in Fig.~\ref{fig:VH}. For this reason, being the $Z$ boson  typically on shell, the $\LO_3$ contribution is not expected to be suppressed w.r.t.~the $\LO_1$ one by a factor of order $(\alpha/\alpha_s)^2$, as one may guess from a naive $\alphas$ and $\alpha$ power counting. On the other hand, $\LO_1$ and $\LO_3$ have completely different shapes at the differential level. The $\LO_2$ instead receives contributions only from  $g\gamma\TO \Hbb$ diagrams (with a ``genuine'' $\Hbb$ topology) and therefore it is expected to be negligible in comparison to the $\LO_1$ and the $\LO_3$. 

Moving to NLO, we do not list here all the possible partonic initial states, but nonetheless we comment on the topologies appearing at any $\NLO_{i}$ order.
On the one hand, $\NLO_1$  and $\NLO_2$ can be viewed as the NLO QCD and NLO EW corrections to the ``genuine'' $\Hbb$ production, respectively. On the other hand,   $\NLO_3$ and $\NLO_4$ can be viewed mainly as the NLO QCD and NLO EW corrections to the $Z(\TO b\bar b)H$ production, respectively. However also new topologies enter the calculation, inducing a sensitivity to new interactions.

$\NLO_1$ receives a contribution from an additional topology: gluon fusion  with an emission of a gluon splitting into a $b \bar b$ pair, $gg$F+$b\bar b$, giving rise to terms of
order $y_b y_t$. Similar contributions are also present in the $\NLO_2$, where the $b \bar b$ pair instead emerges from a photon  or $Z$-boson emission from the loop. At the same order, terms proportional to $y_b y_t$ can be induced also by diagrams such as the one in Fig.~\ref{fig:gg2hbb_yt}, which has a similar topology to the one shown in  Fig.~\ref{fig:ggHbb}, but does not depend on $y_b$. Similarly, diagrams like the one in Fig.~\ref{fig:ggHbb_W} can induce a sensitivity on  the $HWW$ interaction without depending on $y_b$. In general, EW corrections can potentially induce a sensitivity on any other SM electroweak interaction\footnote{Also loop diagrams involving double-Higgs boson production with one Higgs decaying into a $b \bar b$ pair give a contribution to the $\NLO_2$. However, they enter the calculation only via the interference with the tree-level ``genuine'' $\Hbb$ topology, where $b \bar b$ pairs never stem from a resonant Higgs propagator. Therefore, even setting $\Gamma_H=0$, the singularity of such loop diagrams at $m(b \bar b)=m_H$ is integrable.} and in particular, in the case of the Higgs boson, on interactions different from $y_b$,  with much larger coupling constants. 

One of the most important findings of our work is that also the $\NLO_3$ term receives a contribution from an additional new topology, namely, the VBF-like diagrams that appear at this order for the real-emission processes $gq\TO \Hbb q$ (see a representative diagram in Fig.~\ref{fig:VBF}); these diagrams are the 4FS counterpart of $qb\TO H qb$ contributions to VBF in the 5FS. For this reason, $\NLO_3$ contributions are expected to be large and to have different shapes from the $\LO_3$ ones. Similar VBF contributions are present in the  $\NLO_4$, where instead of a gluon, a photon is present in the initial state. Moreover, the argument of the previous paragraph regarding the sensitivity on additional EW interactions introduced by EW corrections applies also to the $\NLO_4$. In this case, the  dominant underlying tree-level topology is the $VH$ one, Fig.~\ref{fig:VH}, but also contributions coming from the $\Hbb$ topology, Fig.~\ref{fig:qqHbb}, with the gluon substituted by a $\gamma/Z$ boson, are present. In conclusion, all the perturbative orders are in principle non-negligible and exhibit different shapes at the differential level.
 
 \begin{table}[!t]
\begin{center}
\begin{tabular}{l|  p{5cm}}
\toprule
Perturbative order & Topologies \\
\midrule
$\LO_1 ~(\alphas^2\alpha)$ & $gg,q\bar q \TO \Hbb$ \\
$\LO_2 ~(\alphas\alpha^2)$ & $\gamma g\TO \Hbb$   \\
$\LO_3 ~(\alpha^3)\phantom{\alpha^2}$ & $q\bar q \TO ZH (Z \TO b \bar b )$,  $ q\bar q, \gamma \gamma\TO \Hbb$\\ 
\midrule
Perturbative order & Topologies \\
\midrule
$\NLO_1 ~(\alphas^3\alpha)$ & $\Hbb$, $\cancel{gg{\rm F}+b \bar b}$\\
$\NLO_2 ~(\alphas^2\alpha^2)$ & $\Hbb$, $gg$F+$b \bar b$ \\ 
$\NLO_3 ~(\alphas\alpha^3)$ & $ZH$, VBF \\
$\NLO_4 ~(\alpha^4)\phantom{\alpha^2}$ & $ZH$, VBF \\
\bottomrule
\end{tabular}
\end{center}
\caption{ Topologies entering at LO, with initial states that are explicitly specified, and at NLO. As discussed in Sec.~\ref{sec:msbar}, the terms proportional to $y_b y_t$ in $\NLO_1$, emerging from the interference of $\Hbb$ and $gg$F+$b\bar b$ topologies, are not taken into account in our calculation.} 
\label{table:orders} 
\end{table}

We remind the reader that the loop-induced $gg$F+$b\bar b$  topology leads to very large contributions to the $\Hbb$ final-state. As discussed in Ref.~\cite{Deutschmann:2018avk}  these contributions involve terms proportional to $y_t^2$ and to $y_b y_t$, where the latter, as already mentioned before, is part of the $\NLO_1$ and originates from the interference of the $gg$F+$b\bar b$  and ``genuine'' $\Hbb$ topologies. The $y_t^2$ term is formally an NNLO QCD effect\footnote{See Sec.~2.1 in Ref.~\cite{Deutschmann:2018avk} for more details.}, but due to the $(y_t/y_b)^2$ enhancement w.r.t.~the ${\rm LO}_1$ it is numerically important, as also shown 
in Sec.~\ref{sec:prospects}. Therefore, the presence of these $y_t$-dependent terms in addition of the very large (reducible) background due to $gg$F plus additional light jets, makes the extraction of  $y_b$ from $\Hbb$ measurement very challenging.
The fact that two additional different topologies ($ZH$ and VBF) are present and give sizeable contributions that depend on the $HZZ$ coupling rather than on $y_b$, 
dramatically complicates the extraction of $y_b$ from $\Hbb$ production at the LHC. As we will quantify better in Sec.~\ref{sec:results}, given the smallness of the $\Hbb$ production cross section already at the inclusive level, possible extra selection cuts that reduce the dependence on $y_t$ and the $HVV$ coupling are not only difficult to design, but also end up in killing
 an already rare signal. 

Before moving to the next section we comment on the reliability of the usage of the pure 4FS in the context of our study.
As we will better explain in Sec.~\ref{sec:msbar},  the two conditions $y_b \ne 0$ and $m_b=0$ are inconsistent when EW corrections are taken into account. Thus, the 4FS remains the only possible choice for performing the computation of complete-NLO predictions to $\Hbb$ production.  However, even considering only QCD corrections, one may argue that in such a scheme the perturbative 
convergence is jeopardised by the presence of large logarithms of the form $\alpha_s^n \log^k(  Q/m_b )$ with $n,k>0$, where $Q$ is a hard scale involved in the process. For $\Hbb$ production, such potential large logarithms are
only of initial-state origin and arise from the initial-state gluon splitting into a $b\bar b$ pair and subsequent gluon emissions from the $b$ quarks, leading to the condition $k\le n$. In our calculation, they appear both in the 
``genuine'' $\Hbb$ topology (twice, once for each  initial-state gluon) as well as in the VBF topology (only once). In the 5FS,  these  logarithms would be resummed  and automatically incorporated in the evolution of the bottom-quark PDF, which  would be the 5FS counterpart of the initial-state gluon splitting into a $b\bar b$ pair that is present in the 4FS diagrams. On the other hand, as it was in general shown in Refs.~\cite{Maltoni:2012pa,Lim:2016wjo},  the impact
of these collinear logarithms is not very large unless the process is dominated by large-$x$ dynamics. It was also shown that the typical scale $Q$ entering the terms $\alpha_s^n \log^k(  Q/m_b )$ in a 4FS calculation is considerably smaller than the naively expected value, {\it i.e.}, the total invariant mass of the final-state (defined excluding the bottom quarks). The same argument also suggests the usage of a rather low value for the factorisation (and renormalisation) scale, as we will do in this work and specify in Sec.~\ref{sec:inputs}. In conclusion, the combination of 4FS and 5FS predictions as done in Refs.~\cite{Forte:2015hba,Forte:2016sja,Bonvini:2015pxa,Bonvini:2016fgf,Duhr:2020kzd} is definitely important for improving the precision of the description of the contribution from the ``genuine'' $\Hbb$ topology, but not compulsory for providing a sensible prediction and the 4FS can be safely used for the purpose of our work. 

A different kind of potentially large logarithms (of the form $ \log^k( \mu_R / \mb)$ where $\mu_R$ is the renormalisation scale) instead arise from the renormalisation of $m_b$ and $y_b$ and have a huge impact, which cannot be neglected in our work, on the predictions for $\Hbb$. We discuss this subject in the next section.

\subsection{Renormalisation of the bottom Yukawa coupling and EW corrections}
\label{sec:msbar}

An important issue in the calculation carried out here is the renormalisation condition for the parameter $m_b$, {\it i.e.}, the mass of the bottom quark.
In our calculation we adopt the complex mass scheme for the massive unstable particles (the $W,Z$ and Higgs bosons and the top quark), the $G_\mu$ scheme for the electroweak interactions and the $\MSbar$ scheme with four active flavours for  $\alphas$. The remaining renormalisation condition is the one concerning $m_b$, which enters the calculation both via the phase-space integration, since there are two on-shell bottom quarks in the final state, and via the matrix elements, especially via the $y_b^2$ dependence.
 The parameter $y_b$ is the Yukawa of the bottom quark and in the SM $y_b=\sqrt{2} m_b/v$,  where $v$ is the Higgs vacuum expectation value. While in our calculation it is natural to set an on-shell condition for $m_b$ itself,  in the case of $y_b$ the typical scale involved in the interaction is much larger, being of the order of the Higgs mass $m_H$. Performing a calculation of higher-order effects of purely  QCD origin, the problem can be solved by treating  $y_b$ and $m_b$ as independent parameters, with $y_b$ renormalised in the $\MSbar$ scheme  at the renormalisation scale $\mu_R$, and $m_b$ on-shell. This strategy has shown an improved convergence of the perturbative series, automatically resumming in $y_b$ large logarithms of the form $\log(\mu_R/m_b)$, and a reduction of the difference with the corresponding calculation in the 5FS.
Moreover, it allows to independently vary $y_b$ and $m_b$.

The problem arises from the fact that when EW corrections are calculated the SM relation $y_b= \sqrt{2} m_b/v$ must be enforced\footnote{This fact is strictly true in the SM. For example, in an effective-field-theory approach the relation $y_b=\sqrt{2} m_b/v$ can be modified while keeping the possibility of performing EW corrections, {\it e.g.},  by adding a dimension-6 operator of the form $\frac{\Phi^\dagger \Phi}{\Lambda^2} \bar Q_L \Phi b_R$, where $\Phi$ is the Higgs doublet.  However, the renormalisation conditions for this kind of calculation is much more involved, see {\it e.g.} Ref.~\cite{Cullen:2019nnr}.}  in order to have the cancellation of UV divergences. This fact has two consequences for a SM calculation at NLO EW or complete-NLO accuracy for $\Hbb$ production: 
\begin{itemize}
\item in the 5FS $y_b= \sqrt{2} m_b/v=0$ and therefore the computation is not feasible,
\item in the 4FS the renormalisation condition for $m_b$ fixes the renormalisation condition for $y_b$ and vice versa.
\end{itemize}

Although the main phenomenological results of this paper are not strictly based on precision physics, namely percent or higher-level accuracy, a correct treatment of the input parameters for the $\LO_1$ and $\NLO_1$ is crucial in order to obtain sensible result; using $y_b=\sqrt{2} \mbpole/v$ can lead to results larger by a factor of 2 to 3 than in the case of $y_b=\sqrt{2} \mbMSbar/v$.

In order to amend this situation, we adopt the following procedure.
First of all, we perform the purely QCD calculation ($\LO_1$ and $\NLO_1$) by employing the $\MSbar$ scheme for the Yukawa of the bottom ($y_b=\sqrt{2} \mbMSbar/v$), and the on-shell scheme for $\mb$. For this reason, it is important that the two input parameters, $\mbMSbar(\mbMSbar)$ and $\mbpole$ respectively, are consistent. Following the recommendation from Sec.~IV.2.2.a of Ref.~\cite{deFlorian:2016spz}, we adopt a one-loop QCD transition between the two schemes.\footnote{The reason is that renormalons are present in the quantity $\mbMSbar(\mbMSbar)-\mbpole$ and the perturbative series does not converge, as discussed in Ref.~\cite{Bonvini:2016fgf}. Therefore, at variance with a general convergent series, the common lore ``the higher is the precision the better it is'' does not apply here. }  At the same time, we exclude the contribution from interference terms between $gg$F+$b\bar b$ and $\Hbb$ topologies in $\NLO_1$, which are proportional to $y_b y_t$ and have already been studied in Ref.~\cite{Deutschmann:2018avk} at higher accuracy and shown to have a mild impact on the cross section.  Second, we combine the calculation of the $\LO_1$ and $\NLO_1$ contributions in this scheme, dubbed $\LO_1^\MSbar$ and $\NLO_1^\MSbar|_{y_t=0}$ respectively, with the remaining complete-NLO terms where the Yukawa of the bottom is renormalised on-shell  $y_b=\sqrt{2} \mbpole/v$.
 In doing so we do not simply add the different perturbative orders, but we combine them in a sort of multiplicative approach, namely, we first define
 
 \begin{equation}
\NLO_2^\MSbar \equiv \NLO_2  \frac{\LO_1^{\MSbar}}{\LO_1} \,, \label{eq:NLO2MS}
\end{equation}
and then
\begin{equation}
(\LO+\NLO)^{\MSbar}\equiv \LO_1^{\MSbar} +\NLO_1^\MSbar|_{y_t=0} + \NLO_2^\MSbar +(\LO_2+\LO_3 +\NLO_3 +\NLO_4)\,. \label{eq:MSandonshell}
\end{equation}
All quantities without a superscript in Eqs.~\eqref{eq:NLO2MS} and \eqref{eq:MSandonshell} are meant with $\mb$ renormalised on shell,  at variance with those with $^{\MSbar}$. 

Beside the renormalisation of the $\Hbb$ vertex, the $\NLO_2$ term contains several other types of contributions, such as EW Sudakov logarithms that depend on $y_b$ only via the underlying $\LO_1$ contribution and therefore can be naturally rescaled by a factor $(\mbMSbar/\mbpole)^2$ in order to provide an improved prediction; the term $\NLO_2^\MSbar$ in Eq.~\eqref{eq:NLO2MS} precisely corresponds to the rescaling of the $\NLO_2$ contribution by this factor. In principle one may think also to add a term $ \NLO_2 \times(\NLO_1^\MSbar|_{y_t=0}/\LO_1)$ as typically done in the multiplicative combination of NLO QCD and EW corrections. We have checked this alternative approach and found negligible differences with the results obtained via Eq.~\eqref{eq:MSandonshell}. Finally, we remind the reader that the orders $\LO_3$,  $\NLO_3$ and $\NLO_4$ are dominated by $ZH$ and VBF configurations, for which the issue concerning the renormalisation of the $\Hbb$ interaction is not relevant.

For simplicity, in the rest of the paper we will use also the notations defined in the following
 \begin{eqnarray}
\LOQCD&\equiv&\LO_1^{\MSbar}\, ,  \label{eq:def1}\\
\LO&\equiv&\LO_1^{\MSbar}+\LO_2+\LO_3\, ,  \label{eq:def2}\\
\NLOQCD&\equiv&\LO_1^{\MSbar} +\NLO_1^\MSbar|_{y_t=0}\, , \label{eq:def3}\\
\NLOQCDEW&\equiv& \LO_1^{\MSbar} +\NLO_1^\MSbar|_{y_t=0} + \NLO_2^\MSbar\, , \label{eq:def4}\\
\CNLO&\equiv&(\LO+\NLO)^{\MSbar}\, . \label{eq:def5}
\end{eqnarray}

\section{Numerical results}

\label{sec:results} 

\subsection{Input parameters}
\label{sec:inputs}
We provide numerical results for proton-proton collisions at the LHC with a center-of-mass-energy of 14 TeV, as planned for the Run-III and the High-Luminosity run \cite{ApollinariG.:2017ojx}.
We perform the calculation using the complex mass scheme and the following input parameters\footnote{Beside $m_b$, the input parameters are the same of Ref.~\cite{Frederix:2018nkq}. See Sec.~4.1 of this reference for more details.} 
\begin{align}
\mz &= 91.15348 ~\gev \, ,&\quad \gamz& = 2.4946 ~\gev \,,& \quad \mw &= 80.35797 ~\gev \,,&\quad \gamw &= 2.08899 ~\gev\,,& \\
\mh &= 125.0  ~\gev\, ,&\quad \gamh&=0\, ,& \quad \mt&= 173.34 ~\gev \,, &\quad \gamt &= 1.3692~\gev\, ,
\end{align}
where we have set $\gamh=0$ since there is an external on-shell Higgs in our calculation.
EW interactions are renormalised in the $G_\mu$-scheme with 
\begin{equation}
    \gmu = 1.16639 \cdot 10^{-5} \gev^{-2}.
\end{equation}
We set the pole mass of the bottom quark to
\begin{equation}
\mbpole = 4.58~\gev\, ,
\end{equation}
which corresponds to 
\begin{equation}
\mbMSbar(\mbMSbar) = 4.18~\gev\, ,
\end{equation}
when the difference between the two schemes is evaluated at one-loop level, as motivated in Sec.~\ref{sec:msbar}. We do not evaluate uncertainties related to the value of $\mb$; they are discussed, {\it e.g.}, in Refs.~\cite{Bonvini:2016fgf,Duhr:2020kzd} and are of the order of a few percents.

For the factorisation scale $\mu_{F}$ and the renormalisation scale $\mu_{R}$, which enters  the definition of $\alpha_s(\mu_{R})$ and $y_b=\sqrt{2} \mbMSbar(\mu_{R})/v$ in the NLO QCD calculation, we use a central value
\begin{equation}
    \mu_{0} = H_T/4, \quad H_T = \sum_i \sqrt{m_i^2 +p^2_T(i)}\, , \label{eq:renscale}
\end{equation}
where the index $i$ runs over all the final-state particles. The $\mu_R$ dependence of $\alpha_s$ is directly taken from the PDF set employed in the calculation, while we evolve $\mbMSbar(\mu_{R})$ at four loop in QCD~\cite{Marquard:2015qpa,Kataev:2015gvt}. Scale uncertainties are evaluated by independently varying the factorisation and renormalisation scale in the range $\mu_0/2<\mu_F,\mu_R<2\mu_0$.

Phase-space integration is performed with no constraints on the $b$-quark momenta. On the other hand, we will provide results for the full phase space as well by setting constraints on the number of $b$-jets and possibly light jets.  Jets are clustered with the
anti-$k_T$ algorithm~\cite{Cacciari:2008gp} as implemented in {\sc \small FastJet}~\cite{Cacciari:2011ma}, with the distance parameter $R=0.4$. Jets are required to have $p_T > 30$ GeV and pseudorapidity $|\eta|<4.5$. Photons are clustered into jets and therefore also a real emission of a single photon can form a separate jet.\footnote{We remind the reader that in many LHC analyses a jet is defined with up to 99\% of its energy of electromagnetic origin and up to 90\% of it that can be carried by a single photon. See Ref.~\cite{Frederix:2016ost} for further details on this subject.} In the case of jets that have at least a bottom quark or antiquark among their constituents, the requirement $|\eta| < 2.5$ has to be satisfied in order to be classified as $b$-jets, otherwise they are tagged as light jets. As we will discuss in Sec.~\ref{sec:inclusive}, we will also explore the effects of a jet-veto in the entire phase-space region $|\eta|<4.5$.

Since we calculate NLO EW corrections, as discussed in Sec.~\ref{sec:topologies}, processes featuring initial-state photons are present and therefore a parameterisation of the photon PDF is necessary. Furthermore NLO QED effects in PDFs evolution have to be taken into account. Even more important, the PDFs should be in the 4FS in order to correctly take into account QCD effects. However, to the best of our knowledge, at the moment there are no public PDF sets including NLO QED effects and a photon density in the 4FS. For this reason, we have checked the numerical impact of the photon PDF and NLO QED effects by comparing results obtained via the usage of the PDFs set \texttt{NNPDF31\_nnlo\_as\_0118\_luxqed}  \cite{Bertone:2017bme}, which is based on the PDF fit  {\sc\small  NNPDF3.1}  \cite{Ball:2017nwa} and the photon parameterisation of {\sc\small LUXqed} \cite{Manohar:2016nzj, Manohar:2017eqh}, and the PDF set \texttt{NNPDF31\_nnlo\_as\_0118}, which includes neither a photon PDF nor NLO QED effects in the evolution. These two PDF sets are both in the 5FS, but this is irrelevant for the sake of estimating QED effects in the PDFs. We find that effects related to the QED evolution are of the order of 1\% for $\LO_1$, while
the  LO$_2$ term, which is purely $\gamma g$-induced, gives a contribution that is 0.1\% of the LO$_1$ one. Therefore, neglecting QED effects in PDFs is completely justified for the study carried out in this work.
We conclude that we can safely use the PDF set \texttt{NNPDF31\_nnlo\_as\_0118\_nf\_4l} which does not account for QED effects, but is designed for calculations in the 4FS.\footnote{In principle, one could use the PDF set \texttt{NNPDF31\_nnlo\_as\_0118\_luxqed} and remove the impact of the fifth
  flavour from the running of
  $\alpha_s$ and to the DGLAP equation for the PDF evolution in order to be consistent at NLO QCD accuracy. For instance, one may adopt a strategy similar to the one explained in Sec.~2.2 of Ref.~\cite{Badger:2016bpw}, which is in turn based on Ref.~\cite{Cacciari:1998it}. However, logarithms of the form
   $\log(\mu_{\rm F,R}/\mbpole)$ would be present and not resummed, especially when varying the scale.} Nevertheless, we advocate the necessity to have public PDF sets including QED effects in the 4FS.

\subsection{Inclusive results}
\label{sec:inclusive}

We now turn to the presentation of results, starting from total cross sections for $\Hbb$ production at 14 TeV defined for different jet categories. In Tab.~\ref{tab:rates}, we list predictions computed at different perturbative accuracies, according to the definitions in Eqs.~\eqref{eq:def1}--\eqref{eq:def5}. We show results for different selection cuts on $b$-jet multiplicities, namely,
\begin{itemize}
    \item{NO CUT: No restriction on the momenta of the final-state particles,}
    \item{ $N_{j_b} = 1$: Exactly one $b$-jet, with and without a veto on light jets,}
    \item{ $N_{j_b} \ge 1$: At least one $b$-jet, with and without a veto on light jets,}
    \item{ $N_{j_b} \ge 2$: At least two $b$-jets.}
\end{itemize} 
At our accuracy, complete-NLO, there cannot be more than two $b$-jets and therefore $N_{j_b} \ge 2 \Longleftrightarrow N_{j_b} = 2$.
Numbers in parentheses refer to the case where the light-jet veto is applied. In the second column we show total rates for the central scales $\mu_F=\mu_R=\mu_0$, together with relative scale uncertainties, while in the third column we show the corresponding ratios with the central-scale $\LOQCD$ predictions. The relative impact of the different perturbative orders is further documented in Tab.~\ref{tab:breakdown}, where the ratios of all the different contributions entering the complete-NLO predictions ($\CNLO$) divided by $\LOQCD$ are separately displayed, see also Eqs.~\eqref{eq:NLO2MS},\eqref{eq:MSandonshell} and \eqref{eq:def5}. We recall that LO$_2$ is exactly zero since we use a PDF set without a photon density,
and therefore its contribution is not displayed in Tab.~\ref{tab:breakdown}. We also remind the reader that the term $\LO_1$ is equivalent to  $\LOQCD$, but with the Yukawa of the bottom renormalised on-shell,  $y_b=\sqrt{2} \mbpole/v$.

\subsubsection{Description of the results}
\label{sec:description}

We start the discussion of the numerical results by  commenting the numbers in Tabs.~\ref{tab:rates} and \ref{tab:breakdown} obtained without applying the light-jet veto. We will then move to the case with the light-jet veto and finally we will draw our phenomenological conclusions in Sec.~\ref{sec:prospects}: at variance with the naive expectation, the measurement of total rates for $\Hbb$ production is not leading to a direct sensitivity to $y_b$, regardless of the selection cuts that are used.

\paragraph{Results without the light-jet veto}

\begin{table*}
    \centering
    \begin{tabular}{c|c|c|c}
        accuracy $(i)$ & $\sigma_i~[\textrm{fb}]$  & $\sigma_i / \sigma_{\LOQCD}$& cuts   \\
\hline
\hline
            $\LOQCD$   &  $ 297 ^{+ 55.9\%} _{- 34.1\%} $  &   1.00       \\
                  LO   &  $ 399 ^{+ 42.9\%} _{- 26.9\%} $  &   1.34       \\
     NLO$_{\rm QCD}$   &  $ 450 ^{+ 19.2\%} _{- 20.7\%} $  &   1.51      &   NO CUT   \\
  NLO$_{\rm QCD+EW}$   &  $ 442 ^{+ 18.5\%} _{- 20.4\%} $  &   1.49       \\
             $\CNLO$   &  $ 639 ^{+ 14.3\%} _{- 15.6\%} $  &   2.15       \\
\hline
            $\LOQCD$   &  $ 67.2 ^{+ 49.1\%} _{- 30.8\%} $ \, ($ 64.6 ^{+ 49.5\%} _{- 31.1\%} $)  &   1.00 \, ( 1.00)   \\
                  LO   &  $ 154 ^{+ 24.2\%} _{- 16.9\%} $ \, ($ 142 ^{+ 25.2\%} _{- 17.5\%} $)  &   2.29 \, ( 2.19)   \\
     NLO$_{\rm QCD}$   &  $ 94.4 ^{+ 12.3\%} _{- 16.2\%} $ \, ($ 69.6 ^{+  2.3\%} _{- 11.3\%} $)  &   1.40 \, ( 1.08)  &   $N_{j_b} \ge 1$   \\
  NLO$_{\rm QCD+EW}$   &  $ 92.0 ^{+ 11.4\%} _{- 15.8\%} $ \, ($ 67.3 ^{+  2.4\%} _{- 10.6\%} $)  &   1.37 \, ( 1.04)   \\
             $\CNLO$   &  $ 247 ^{+  8.9\%} _{-  8.9\%} $ \, ($ 139 ^{+  0.9\%} _{-  5.3\%} $)  &   3.67 \, ( 2.15)   \\
\hline
            $\LOQCD$   &  $ 61.7 ^{+ 49.6\%} _{- 31.1\%} $ \, ($ 59.0 ^{+ 50.0\%} _{- 31.3\%} $)  &   1.00 \, ( 1.00)   \\
                  LO   &  $ 105 ^{+ 31.1\%} _{- 20.8\%} $ \, ($ 93.3 ^{+ 33.7\%} _{- 22.3\%} $)  &   1.71 \, ( 1.58)   \\
     NLO$_{\rm QCD}$   &  $ 87.9 ^{+ 13.1\%} _{- 16.6\%} $ \, ($ 66.0 ^{+  2.2\%} _{- 12.3\%} $)  &   1.43 \, ( 1.12)  &   $N_{j_b} = 1$   \\
  NLO$_{\rm QCD+EW}$   &  $ 85.7 ^{+ 12.2\%} _{- 16.3\%} $ \, ($ 63.9 ^{+  2.3\%} _{- 11.7\%} $)  &   1.39 \, ( 1.08)   \\
             $\CNLO$   &  $ 187 ^{+ 10.4\%} _{- 10.6\%} $ \, ($ 107 ^{+  1.3\%} _{-  8.4\%} $)  &   3.03 \, ( 1.82)   \\
\hline
            $\LOQCD$   &  $  5.57 ^{+ 45.4\%} _{- 29.0\%} $  &   1.00       \\
                  LO   &  $ 48.4 ^{+  9.0\%} _{-  8.2\%} $  &   8.70       \\
     NLO$_{\rm QCD}$   &  $  6.53 ^{+  1.8\%} _{- 10.8\%} $  &   1.17      &   $N_{j_b} \ge 2$   \\
  NLO$_{\rm QCD+EW}$   &  $  6.30 ^{+  1.0\%} _{- 10.2\%} $  &   1.13       \\
             $\CNLO$   &  $ 59.8 ^{+  4.0\%} _{-  3.7\%} $  &  10.75       \\
    \end{tabular}
    \caption{\label{tab:rates}  Cross sections, with relative scale uncertainties, at different perturbative accuracies and with different phase-space cuts. Numbers in
parentheses are obtained by vetoing light jets. Details are explained in the text.}
\end{table*}

\setlength{\tabcolsep}{4pt}

\begin{table*}
\small
    \centering
    \begin{tabular}{c|c|c|c|c|c|cccc}

$\sigma_i / \sigma_{\LOQCD}$ [\%]    &  LO$_1$  &     LO$_3$  &   NLO$_1^{\MSbar}|_{y_t=0}$  &    NLO$_2^{\MSbar}$  &    NLO$_3$  &   NLO$_4$  &       \\
\hline
\hline
               NO CUT   &  219.1      &   34.1      &   51.3      &   -2.6      &   34.6      &   -2.5       \\
\hline
      $N_{j_b} \ge 1$   &  229.5 \, (229.2)  &  128.7 \, (119.5)  &   40.5 \, (  7.9)  &   -3.6 \, ( -3.6)  &  111.1 \, (  0.9)  &   -9.6 \, ( -9.3)   \\
\hline
        $N_{j_b} = 1$   &  228.6 \, (228.1)  &   70.8 \, ( 58.1)  &   42.5 \, ( 11.9)  &   -3.5 \, ( -3.5)  &   98.7 \, ( 20.0)  &   -5.3 \, ( -4.6)   \\
\hline
      $N_{j_b} \ge 2$   &  240.5      &  770.2      &   17.3      &   -4.1      &  248.4      &  -56.7           \end{tabular}
    \caption{\label{tab:breakdown}  Ratio with the $\LOQCD$ contribution for the $\LO_1$ prediction and for each perturbative order entering the complete-NLO predictions ($\CNLO$).  Numbers are in percentage and those in
parentheses are obtained by vetoing light jets. Details are explained in the text. }
\end{table*}

\setlength{\tabcolsep}{10pt}

The most important feature that can can be observed in Tab.~\ref{tab:breakdown} is that the relative impact of  $\LO_3$,  $\NLO_3$ and $\NLO_4$ grows with $N_{j_b}$. First of all, these contributions, before setting cuts, are not dominated  by the ``genuine'' $\Hbb$ topology, but rather by the $ZH$ and (except $\LO_3$) VBF topologies. Then, while in the  ``genuine'' $\Hbb$ topology with the gluon-gluon initial state  (Fig.~\ref{fig:ggHbb}), which dominates $\LOQCD$, NLO$_1^{\MSbar}$  and    NLO$_2^{\MSbar}$,    both the bottom quarks tend to be collinear to the beam-pipe axis, in the VBF topology this holds true for only one of the two bottom quarks and for none of them in $VH$.  Therefore, the probability that a bottom quark $b$ is either soft or falls outside the rapidity region in which $b$-jets are tagged, $|\eta(j_b)|<2.5$, is higher for  the ``genuine'' $\Hbb$ topology than for $ZH$ and VBF topologies. The same behaviour has been observed in Ref.~\cite{Deutschmann:2018avk} regarding the comparison with the $gg$F+$b \bar b$ topology. The net effects is the aforementioned growth  of the relative impact of  $\LO_3$,  $\NLO_3$ and $\NLO_4$ with $N_{j_b}$.

The same feature can be observed also in  Tab.~\ref{tab:rates} by comparing the $\LOQCD$, $\NLOQCD$ and $\NLOQCDEW$ predictions, which do not include  the  $\LO_3$,  $\NLO_3$ and $\NLO_4$ contributions, with the $\LO$ and $\CNLO$ ones, which do include (part of) them. In fact, according to  Eqs.~\eqref{eq:def1}--\eqref{eq:def5}, since we set the photon PDF to zero, we exactly have $\LO=\LOQCD+\LO_3$ and $\CNLO=\NLOQCDEW+\LO_3+\NLO_3+\NLO_4$.  Therefore,  as already demonstrated in Refs.~\cite{Biedermann:2016yds, Biedermann:2017bss, Frederix:2017wme, Frederix:2019ubd} for other processes, contributions formally suppressed by the $(\alpha/\alpha_s)$ naive power counting can actually be  numerically much larger than expected, especially when specific phase-space cuts are imposed. We remind the reader that  each of the rates for $N_{j_b} \ge 1$ in  Tab.~\ref{tab:rates} is equal to the sum of the corresponding ones for  $N_{j_b} =1$ and $N_{j_b}\ge2$. By looking at the numbers for $N_{j_b}\ge2$ one can understand the large difference between the case $N_{j_b} =1$ and $N_{j_b} \ge 1$. With $N_{j_b}\ge2$ the complete-NLO prediction, $\CNLO$, is 10.8 times larger than the $\LOQCD$ one. The $\LO_3$ is 7.9 times larger than the $\LOQCD$, the $\NLO_3$ is 2.4 times larger, and the $\NLO_4$ is -60\% of the $\LOQCD$. As an example, via a naive $(\alpha/\alpha_s)$ power counting the $\NLO_4$ would be expected to be of the order of 0.01\% of the $\LOQCD$. Although smaller in size,   a similar pattern is observed also for the case $N_{j_b} =1$ and therefore also for the case $N_{j_b} \ge1$. One can also notice that moving from $N_{j_b}\ge 1$ to $N_{j_b}=1$, the $\LO_3$  contribution is strongly reduced, roughly by a factor of 11, while the $\NLO_3$ is reduced much less, roughly by a factor of 2.5. This is a clear sign that the contribution of the VBF topology to the $\NLO_3$ is sizeable. While the $ZH$ topology tend to have two separate $b$-jets, the VBF one mostly exhibits a bottom-quark collinear to the beam-pipe axis and the other one sufficiently central in order to form a $b$-jet. Therefore, once the $N_{j_b}\ge2$ contribution is removed, only the $ZH$ topology is strongly suppressed. This argument will be corroborated by the analysis of the $m(j_{b,1},j_{b,2})$ distribution, {\it i.e.}, the invariant mass of the two $b$-jets,  which is presented in Sec.~\ref{sec:differential}.

Regarding the NLO$_2^{\MSbar}$ contribution, {\it i.e.} what is typically denoted as the NLO EW corrections, it is of the size expected by the naive $(\alpha/\alpha_s)$ power counting:  of the order of a few percents of $\LOQCD$ predictions. Moreover, it mildly depends on the value of $N_{j_b}$. The reason is that at this order there are no new topologies opening, at variance with the $\NLO_3$ and $\NLO_4$ cases. If we had  not consider the quantity NLO$_2^{\MSbar}$, as defined in Eq.~\eqref{eq:NLO2MS} (see also Eq.~\eqref{eq:def1}), but directly $\NLO_2$ from Eq.~\eqref{eq:blobNLO}, the contribution of NLO EW corrections would have been larger. Indeed, as can be seen in Tab.~\ref{tab:breakdown}, the ratio  $(\LOQCD/\LO_1)$ is $\sim$2.4. This ratio has a small dependence on $N_{j_b}$ that is induced by the renormalisation scale of $y_b$, which is dynamical (see Eq.~\eqref{eq:renscale}) and therefore induces not only a global rescaling w.r.t.~the $\LO_1$ term, which has been calculated with on-shell $y_b$, but also  mild differences in shapes. As already mentioned,  NLO EW corrections have already been calculated in Ref.~\cite{Zhang:2017mdz}. However, at variance with Ref.~\cite{Zhang:2017mdz}, not only we identify NLO EW corrections as the NLO$_2^{\MSbar}$ term rather than simply the $\NLO_2$ one, but we also include all the possible initial states contributing to this order. In Ref.~\cite{Zhang:2017mdz}, only the gluon-gluon initial state has been considered.       

NLO QCD corrections, namely the $\NLO_1^\MSbar|_{y_t=0}$ term, have already been calculated in the past \cite{Dittmaier:2003ej,Dawson:2003kb} and are sizeable. Still, with the exception of the case ``NO CUT'', they are in general smaller than the $\NLO_3$ and $\LO_3$ contributions. On the other hand, the $\NLO_1^\MSbar|_{y_t=0}$ term is especially relevant for what concerns scale uncertainties. While the $\LOQCD$ predictions have relative scale uncertainties of the order $\sim^{+ 50\%} _{- 30\%} $, $\NLOQCD$ predictions have relative scale  uncertainties of the order 15-20\% and even smaller for the $N_{j_b}\ge2$ case. If we had not  implemented the $\MSbar$ scheme for  $y_b$, scale uncertainties would had been smaller at LO in QCD ($\LO_1$), since $y_b$ would not depend on $\mu_R$, and also at NLO in QCD ($\LO_1+\NLO_1|_{y_t=0}$).  However, this reduction of scale uncertainties should be interpreted as an underestimate of higher-order effects by the use of $y_b$ in the on-shell scheme rather than a more accurate prediction.  Concerning the $\NLO_2^\MSbar$ term, its  impact  on scale uncertainties is below the 1$\%$ level, as it can be seen by comparing $\NLOQCD$ and $\NLOQCDEW$ predictions. Instead, moving from $\NLOQCDEW$ to $\CNLO$ predictions, the  size of the scale-uncertainty band decreases in any $N_{j_b}$ category. The reason is that the $\LO_3$ contribution has a much smaller scale dependence w.r.t.~the $\LOQCD$ one, since at this order the $ZH$ topology does not depend neither on $y_b$ nor on $\alpha_s$; its scale dependence originates only from PDFs and thus from $\mu_F$. This can be seen by comparing the $\LOQCD$ predictions with the $\LO$ ones, where the latter are exactly equal to the former plus the $\LO_3$ contribution. The $\NLO_3$ contribution introduces a $\mu_R$ dependence via the presence of one power of $\alpha_s$,  but it also further reduces the dependence on $\mu_F$. Altogether, these effects lead to the reduction of the size of the scale-uncertainty band from $\NLOQCDEW$ to $\CNLO$.

\paragraph{Results including the light-jet veto}

We now comment the results where the veto on light jets is applied, namely, the number of Tabs.~\ref{tab:rates} and \ref{tab:breakdown} that are in parentheses. First of all, it is worth to notice that the light-jet veto affects also $\LO_i$ results because $b$-jets are tagged only in the $|\eta(j_b)|<2.5$ region. When $2.5<|\eta(j_b)|<4.5$ the jet is actually tagged as light and therefore the light-jet veto has an effect on it. 
Moving to $\NLOQCD$, $\NLOQCDEW$, and $\CNLO$ predictions, the first comment about them is that scale uncertainties for results with the jet veto do not largely increase w.r.t.~the corresponding cases without it, rather they mildly decrease. This is a clear sign that jet-veto resummation or the matching with the shower effects is not mandatory for obtaining sensible results. The situation is slightly different in the case $N_{j_b}\ge2$, where we have observed much larger scale uncertainties  and therefore we have omitted them in Tabs.~\ref{tab:rates} and \ref{tab:breakdown}. The case ``NO CUT'', without the light-jet veto, has been reported in Tabs.~\ref{tab:rates} and \ref{tab:breakdown} in order to document the result of our calculation and better interpret the $N_{j_b}$ categorisation. On the other hand, we already know that its contribution is about 100 times smaller than inclusive $gg$F production and therefore not suitable for a sensitivity-study on $\Hbb$ production and especially on $y_b$. For this reason, we have chosen to not show the case of a light-jet veto and $N_{j_b}\ge 0$, and in conclusion we consider the light-jet veto option only for the cases $N_{j_b}=1$ and $N_{j_b}\ge1$.

Like in any fixed-order calculation, the light-jet veto has a sizeable impact on the NLO QCD $K$-factor, {\it i.e.}, the $\sigma_{\NLOQCD}/\sigma_{\LOQCD}$ ratio, as can be seen in Tab.~\ref{tab:rates}. Non-negligible effects are present also for the $\LO_3$ and therefore the $\LO$ predictions, as can be respectively seen in Tabs.~\ref{tab:breakdown} and \ref{tab:rates}. However, the largest impact of the light-jet veto is on the $\NLO_3$ contribution and therefore the $\CNLO$ predictions. While without the light-jet veto the $\NLO_3$ contribution is of the same size of the $\LOQCD$ one for both the $N_{j_b}\ge 1$ and $N_{j_b}= 1$ categories, applying the light-jet veto the (central value of the)  $\NLO_3$ contribution almost vanishes  in the case of $N_{j_b}\ge 1$ and drops  to only $\sim20\%$ of the $\LOQCD$ one when $N_{j_b}=1$. The reason is that the VBF topology typically has one light-jet induced by the light quark in the final state and possibly one additional light-jet due to one of the two bottom quarks, which is usually  at large rapidities. Therefore the veto has a huge effect on the contribution from this topology. Moreover, the $\NLO_3$ has a large contribution from ``QCD corrections'' to the $ZH$ topology, which includes gluon emissions from the bottom quarks from the $Z$ decays. The light-jet veto has a large impact also on these configurations, especially in the case of  $N_{j_b}= 2$, which is present also in  $N_{j_b}\ge 1$. This is the reason why the effect of the light-jet veto on the $\NLO_3$  contributions is slightly larger in the case $N_{j_b}\ge 1$ than in the case $N_{j_b}=1$. As a last remark, we notice that the impact of the light-jet veto is instead negligible on NLO$_2^{\MSbar}$ and $\NLO_4$ contributions.

\subsubsection{Prospects on the $y_b$ measurement}
\label{sec:prospects}

On the basis of the previous discussion and of the results of Tabs.~\ref{tab:rates} and \ref{tab:breakdown}, we now comment on what are the prospects of a direct determination of $y_b$ via the $\Hbb$ measurement at the LHC. For the sake of clarity, in the following discussion we will associate specific perturbative orders to specific Higgs couplings:
 \begin{eqnarray}
\LOQCD&\Longrightarrow&\ord(y_b^2)\, , \label{eq:rel1}\\
\NLO_1^\MSbar|_{y_t=0}& \Longrightarrow&\ord(y_b^2)\, , \label{eq:rel2}\\
\NLO_2^\MSbar& \Longrightarrow&\ord(y_b^2)\, , \label{eq:rel3}\\
\LO_3&\Longrightarrow&\ord(\kappa_Z^2) \, ,\label{eq:rel4}\\
\NLO_3&\Longrightarrow&\ord(\kappa_Z^2)\, , \label{eq:rel5}\\
\NLO_4&\Longrightarrow&\ord(\kappa_Z^2)\, , \label{eq:rel6}
\end{eqnarray}
where adopting the $\kappa$-framework  notation \cite{Heinemeyer:2013tqa} we denote the $HZZ$ interaction as $\kappa_Z$. Relations \eqref{eq:rel1}--\eqref{eq:rel6}  also imply
 \begin{eqnarray}
 \NLOQCD&\Longrightarrow&\ord(y_b^2)\, , \label{eq:rel7}\\
\NLOQCDEW&\Longrightarrow&\ord(y_b^2)\, , \label{eq:rel8}\\
\CNLO-\NLOQCDEW&\Longrightarrow&\ord(\kappa_Z^2)\, . \label{eq:rel8}
\end{eqnarray}

Clearly, as also pointed out in Sec.~\ref{sec:topologies}, the $\NLO_2^\MSbar$ and $\NLO_4$ terms involve contributions that depend on additional couplings and that can even not depend at all on $y_b$ and $\kappa_Z$, respectively. However, one can understand from the discussion of Sec.~\ref{sec:description} that the numerical impact of $\NLO_2^\MSbar$ and $\NLO_4$ terms, and therefore of such contributions, is negligible w.r.t.~the other perturbative orders involved in the calculation. Moreover, as it will become more clear in the following, taking into account a more realistic and more complex coupling structure in a given perturbative order would make our argument even stronger. In other words, relations  \eqref{eq:rel1}--\eqref{eq:rel8} are devised for simplifying the discussion, but our conclusions do not depend on them.

\begin{table*}
\small
    \centering
    \begin{tabular}{c|c|c|ccccccc}

ratios    &  $ \frac{\sigma(y_b^2)}{\sigma(y_b^2) + \sigma(\kappa_Z^2)}\equiv \frac{\sigma_{\NLOQCDEW }}{ \sigma_{\CNLO}}  $&  $\frac{\sigma(y_b^2)}{\sigma(y_b^2)+\sigma(y_t^2)+\sigma(y_b y_t) }$   &  $\frac{\sigma(y_b^2)}{\sigma(y_b^2)+\sigma(y_t^2)+\sigma(y_b y_t) +\sigma(\kappa_Z^2)}$         \\
&($y_b$ vs.~$\kappa_Z$) & ($y_b$ vs.~$y_t$) & ($y_b$ vs.~$\kappa_Z$ and $y_t$)\\
\hline
\hline
NO CUT & 0.69   &     0.32     &   0.28            \\
\hline
$N_{j_b} \ge 1$      &  0.37 \, (0.48)     &  0.19  &   0.14  \\
\hline
$N_{j_b} = 1$    &  0.46 \, (0.60)      &  0.20   &   0.16  \\
\hline
$N_{j_b} \ge 2$      & 0.11   &  0.11     &   0.06         \\

    \end{tabular}
    \caption{\label{tab:couplings} Fraction of the cross section scaling as $y_b^2$ for different phase-space cuts. The first column is based on the results from our calculation in Tab.~\ref{tab:rates}. The second column is based on results from Ref.~\cite{Deutschmann:2018avk}. The third column is based on the numbers in the first and second column.  Details are explained in the text.}
\end{table*}

For the same $N_{j_b}$ categories of Tabs.~\ref{tab:rates} and \ref{tab:breakdown}, in the first column of Tab.~\ref{tab:couplings} we report the ratio of the   $\NLOQCDEW$ and $\CNLO$ predictions, here denoted as  $\sigma_{\NLOQCDEW}$ and $\sigma_{\CNLO}$. Both of them are  our best predictions for respectively  the $\ord(y_b^2)$ cross section, denoted in the following also as $\sigma(y_b^2)$, and the sum of it with the  $\ord(\kappa_Z^2)$ cross section, denoted in the following also as $\sigma(\kappa_Z^2)$. Via the ratio $\sigma_{\NLOQCDEW }/ \sigma_{\CNLO}$ we can determine the fraction of the measured cross section that actually depends on $y_b$. Once again, we remind the reader that the case ``NO CUT'' is purely academic, since the signal from inclusive $gg$F Higgs production exceeds the one of $\Hbb$ production by a factor of 100. Thus, one needs to tag at least one $b$-jet and we already know that also after that the $gg$F+$b\bar b$ contribution is large, so we should at least suppress the $ZH$ and VBF topologies, which yield $\sigma(\kappa_Z^2)$. The category $N_{j_b}\ge2$ has very small rates (see Tab.~\ref{tab:rates}) and the lowest  $\sigma_{\NLOQCDEW }/ \sigma_{\CNLO}$ ratio, due to the large contribution of the $ZH$ topology,  therefore it is not expected to be the best option in order to gain sensitivity on $y_b$. This also explains why the category $N_{j_b}=1$, which does not include $N_{j_b}\ge2$, has a larger $\sigma_{\NLOQCDEW }/ \sigma_{\CNLO}$ ratio w.r.t.~the category $N_{j_b}\ge1$, which does include it. However, in both the $N_{j_b}=1$ and $N_{j_b}\ge1$ categories, the VBF contribution is still large, but the light jet-veto (numbers in parentheses) helps in reducing it. In conclusion, the best option seems to be the $N_{j_b}=1$ category with a light-jet veto, where 60\% of the signal depends on $y_b$. 

So far, however, we have completely neglected the contribution of the $gg$F+$b \bar b$ topology, which leads to $\ord(y_b y_t)$ contributions, $\sigma(y_b y_t)$, and especially $\ord(y_t^2)$ contributions, $\sigma(y_t^2)$. In order to amend this situation we use the results of Ref.~\cite{Deutschmann:2018avk}, where $\sigma(y_b^2)$, $\sigma(y_b y_t)$, and $\sigma(y_t^2)$ have been calculated at NLO QCD accuracy. Using the numbers of Tab.~1 in Ref.~\cite{Deutschmann:2018avk}, in the second column of Tab.~\ref{tab:couplings} we report the ratio of the cross section calculated including only the ``genuine'' $\Hbb$ topologies or adding also  the $gg$F+$b \bar b$ one. In other words,  $\sigma(y_b^2)$ divided by $\sigma(y_b^2)$+$\sigma(y_b y_t)$+$\sigma(y_t^2)$. As can be seen, the impact of $\sigma(y_b y_t)$ and $\sigma(y_t^2)$ is huge and therefore cannot be neglected for our purposes.

The same definitions of $b$-jets have been used in Ref.~\cite{Deutschmann:2018avk} and in our work. A few differences in the input parameters are present, but their impact is expect to be minor, especially when ratios of cross sections are considered. In particular, we have explicitly checked that the difference for the collision energy, 13 TeV in Ref.~\cite{Deutschmann:2018avk} and 14 TeV in the present work, has little effect on the ratios. The only results that we cannot derive from Ref.~\cite{Deutschmann:2018avk} are those for the case with a light-jet veto.
On the other hand, for the case without a light-jet veto, we can combine the results from the first and second column of Tab.~\ref{tab:couplings}. Since in the second column we have $\sigma(y_b^2)$ divided by $\sigma(y_b^2)$+$\sigma(y_b y_t)$+$\sigma(y_t^2)$, if we assume that the first column is $\sigma(y_b^2)$ divided by $\sigma(y_b^2)$+$\sigma(\kappa_Z^2)$, we can derive $\sigma(y_b^2)$ divided by $\sigma(y_b^2)$+$\sigma(\kappa_Z^2)$+$\sigma(y_b y_t)$+$\sigma(y_t^2)$, which is the quantity displayed in the third column. The result is striking: in none of the realistic $N_{j_b}$ categories $\sigma(y_b^2)$, {\it i.e.}, the component of the total cross section that scales as $y_b^2$, is larger than $16\%$. As we will see in the next section, differential information is also in general not helping in improving this ratio. Also, the light-jet veto option cannot substantially alter this picture, as can be seen by the number in the first column of Tab.~\ref{tab:couplings}, so this option can also be safely ruled out.

We want to stress that, if we consider $\sigma(y_b^2)$ as the ``signal'' in an experimental analysis, in this work we are not considering a realistic comparison between the signal and its backgrounds. At this stage, regarding the backgrounds, we are considering only the irreducible  backgrounds, without even taking into account the Higgs boson decays. Needless to say, if we had taken into account also the irreducible and reducible backgrounds for a given signature that is induced by a specific Higgs-boson decay, the situation could have only got worse. From the theoretical side, the same applies if instead of assuming the simplified relations  \eqref{eq:rel1}--\eqref{eq:rel8} we would have taken into account the complete coupling dependence. In the next section, we will explore the last hopes of identifying phase-space regions where the sensitivity on $\sigma(y_b^2)$ may be strongly enhanced. We can anticipate, that this is not the case.\\

\subsection{Differential distributions}
\label{sec:differential}

We start the discussion about differential distributions by analysing the $m(j_{b,1},j_{b,2})$ observable, which can be obtained in our analysis only for $N_{j_b}\ge1$ and $N_{j_b}\ge2$ and is exactly the same in the two cases, since $m(j_{b,1},j_{b,2})$ is defined only for $N_{j_b}=2$. By looking at this distribution we can definitely prove that the $\NLO_3$ order is populated by large contributions from the VBF topology, beside the $ZH$ one. After that, we will consider many more observables for the cases $N_{j_b}\ge1$ and $N_{j_b}=1$.

In Fig.~\ref{fig:mbb}, we show the $m(j_{b,1},j_{b,2})$ distribution at different accuracies ($\LOQCD$, $\LO$, $\NLOQCD$, $\NLOQCDEW$, $\CNLO$) together with their scale uncertainties. The left plot refers to the case where the light-jet veto has not been applied, while in the right one we show results with the light-jet veto. In each plot, we show in the lower inset the same quantities of the main panel normalised to the central value of the $\NLOQCD$ prediction. 

As can be seen in  Fig.~\ref{fig:mbb}, the  $m(j_{b,1},j_{b,2})\sim m_Z$ region is completely dominated by the $\LO$ prediction, which contains the $\LO_3$ contribution, the one involving the $ZH$ topology. The $\NLO_3$ contribution, which is contained in the $\CNLO$ prediction, involves QCD corrections to the $ZH$ topology, such as the emission of gluons from the $b \bar b$ pair stemming from the $Z$ boson decay. The radiation of gluons form the $b$ quarks together with  the presence of the $Z$ resonance leads to a large amount of events migrating from the $m(j_{b,1},j_{b,2})\sim m_Z$ region to smaller values of $m(j_{b,1},j_{b,2})$. This behaviour is typical for any invariant mass distribution of decay products of a resonance, when either QCD or QED emissions are considered. However, at variance with this standard picture, in the left plot of Fig.~\ref{fig:mbb} we can see that the difference between the $\CNLO$ and $\LO$ prediction, which is mainly induced by the $\NLO_3$ contribution, is large also for $m(j_{b,1},j_{b,2})\gg m_Z$. This effect is precisely induced by the presence of  VBF configurations, which on the other hand are suppressed when a light-jet veto is applied, as can be seen in the right plot. In Tabs.~\ref{tab:rates} and \ref{tab:breakdown}, we did not show results with the light-jet veto for $N_{j_b}\ge 2$ since scale uncertainties are too large. Indeed, this feature can be seen in the right plot. The analysis of the $m(j_{b,1},j_{b,2})$ spectrum shows also that even applying a cut around the $m(j_{b,1},j_{b,2})=m_Z$ value, the result would be still contaminated by VBF configurations.

\begin{figure}[t!]
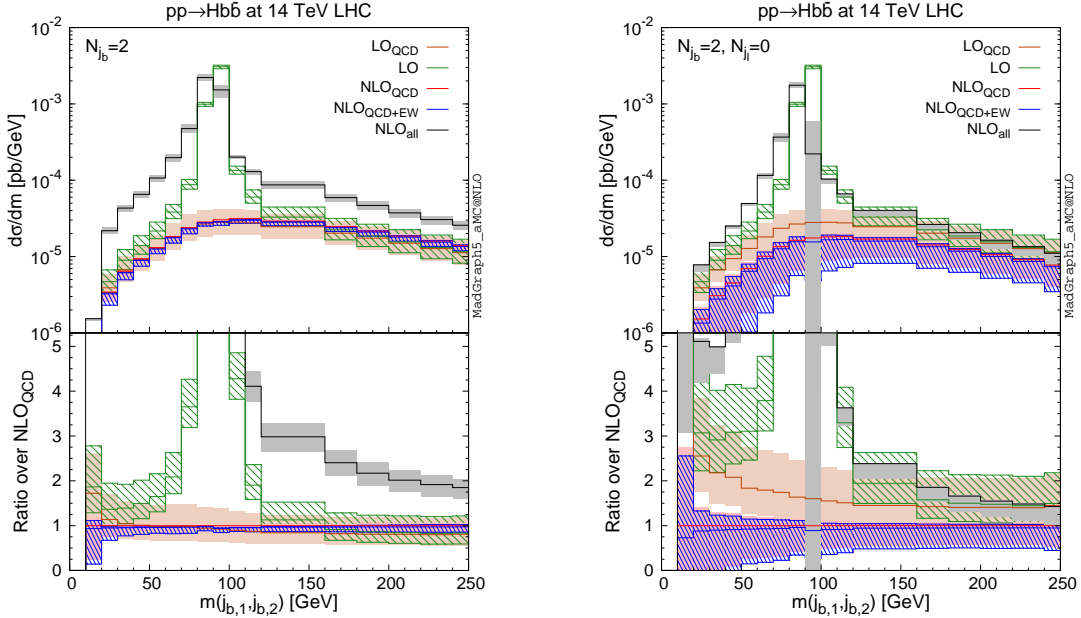

\centering
\foreach \page in {1}{
\includegraphics[page=\numexpr \page\relax, width=.45\textwidth,draft=false]{pp_Hbb_GE2BJ_MSbar2.pdf}
\includegraphics[page=\numexpr \page\relax, width=.45\textwidth,draft=false]{pp_Hbb_GE2BJ0L_MSbar2.pdf}
}
\\
\vspace*{-10mm}
\caption{The $m(j_{b,1},j_{b,2})$ distribution for $N_{j_b}\ge2$. In the right plot the light-jet veto is applied.}\label{fig:mbb}
\end{figure}

\begin{figure}[t!]
\centering
\foreach \page in {1}{
\includegraphics[page=\numexpr \page\relax, width=.45\textwidth,draft=false]{pp_Hbb_GE1BJ_MSbar2.pdf}
\includegraphics[page=\numexpr \page\relax, width=.45\textwidth,draft=false]{pp_Hbb_GE1BJ0L_MSbar2.pdf}\\
\vspace*{-10mm}
\includegraphics[page=\numexpr \page\relax, width=.45\textwidth,draft=false]{pp_Hbb_EQ1BJ_MSbar2.pdf}
\includegraphics[page=\numexpr \page\relax, width=.45\textwidth,draft=false]{pp_Hbb_EQ1BJ0L_MSbar2.pdf}\\
\vspace*{-10mm}
}
\caption{The $p_T(j_{b,1})$ distribution for $N_{j_b}\ge1$ (up) and $N_{j_b}=1$ (down). In the right plots the light-jet veto is applied.}\label{fig:ptjb1}
\end{figure}

\begin{figure}[t!]
\centering
\foreach \page in {4}{
\includegraphics[page=\numexpr \page\relax, width=.45\textwidth,draft=false]{pp_Hbb_GE1BJ_MSbar2.pdf}
\includegraphics[page=\numexpr \page\relax, width=.45\textwidth,draft=false]{pp_Hbb_GE1BJ0L_MSbar2.pdf}\\
\vspace*{-10mm}
\includegraphics[page=\numexpr \page\relax, width=.45\textwidth,draft=false]{pp_Hbb_EQ1BJ_MSbar2.pdf}
\includegraphics[page=\numexpr \page\relax, width=.45\textwidth,draft=false]{pp_Hbb_EQ1BJ0L_MSbar2.pdf}\\
\vspace*{-10mm}
}
\caption{The $\eta(j_{b,1})$ distribution for $N_{j_b}\ge1$ (up) and $N_{j_b}=1$ (down). In the right plots the light-jet veto is applied.}\label{fig:etajb1}
\end{figure}

\begin{figure}[t!]
\centering
\foreach \page in {2}{
\includegraphics[page=\numexpr \page\relax, width=.45\textwidth,draft=false]{pp_Hbb_GE1BJ_MSbar2.pdf}
\includegraphics[page=\numexpr \page\relax, width=.45\textwidth,draft=false]{pp_Hbb_GE1BJ0L_MSbar2.pdf}\\
\vspace*{-10mm}
\includegraphics[page=\numexpr \page\relax, width=.45\textwidth,draft=false]{pp_Hbb_EQ1BJ_MSbar2.pdf}
\includegraphics[page=\numexpr \page\relax, width=.45\textwidth,draft=false]{pp_Hbb_EQ1BJ0L_MSbar2.pdf}\\
\vspace*{-10mm}
}
\caption{The $p_T(H)$ distribution for $N_{j_b}\ge1$ (up) and $N_{j_b}=1$ (down). In the right plots the light-jet veto is applied.}\label{fig:ptH}
\end{figure}

\begin{figure}[t!]
\centering
\foreach \page in {3}{
\includegraphics[page=\numexpr \page\relax, width=.45\textwidth,draft=false]{pp_Hbb_GE1BJ_MSbar2.pdf}
\includegraphics[page=\numexpr \page\relax, width=.45\textwidth,draft=false]{pp_Hbb_GE1BJ0L_MSbar2.pdf}\\
\vspace*{-10mm}
\includegraphics[page=\numexpr \page\relax, width=.45\textwidth,draft=false]{pp_Hbb_EQ1BJ_MSbar2.pdf}
\includegraphics[page=\numexpr \page\relax, width=.45\textwidth,draft=false]{pp_Hbb_EQ1BJ0L_MSbar2.pdf}\\
\vspace*{-10mm}
}
\caption{The $y(H)$ distribution for $N_{j_b}\ge1$ (up) and $N_{j_b}=1$ (down). In the right plots the light-jet veto is applied.}\label{fig:yH}
\end{figure}

\begin{figure}[t!]
\centering
\foreach \page in {5}{
\includegraphics[page=\numexpr \page\relax, width=.45\textwidth,draft=false]{pp_Hbb_GE1BJ_MSbar2.pdf}
\includegraphics[page=\numexpr \page\relax, width=.45\textwidth,draft=false]{pp_Hbb_GE1BJ0L_MSbar2.pdf}\\
\vspace*{-10mm}
\includegraphics[page=\numexpr \page\relax, width=.45\textwidth,draft=false]{pp_Hbb_EQ1BJ_MSbar2.pdf}
\includegraphics[page=\numexpr \page\relax, width=.45\textwidth,draft=false]{pp_Hbb_EQ1BJ0L_MSbar2.pdf}\\
\vspace*{-10mm}
}
\caption{The $|\Delta\eta(H,j_{b,1})|$ distribution for $N_{j_b}\ge1$ (up) and $N_{j_b}=1$ (down). In the right plots the light-jet veto is applied.}\label{fig:DHjb1}
\end{figure}

\begin{figure}[t!]
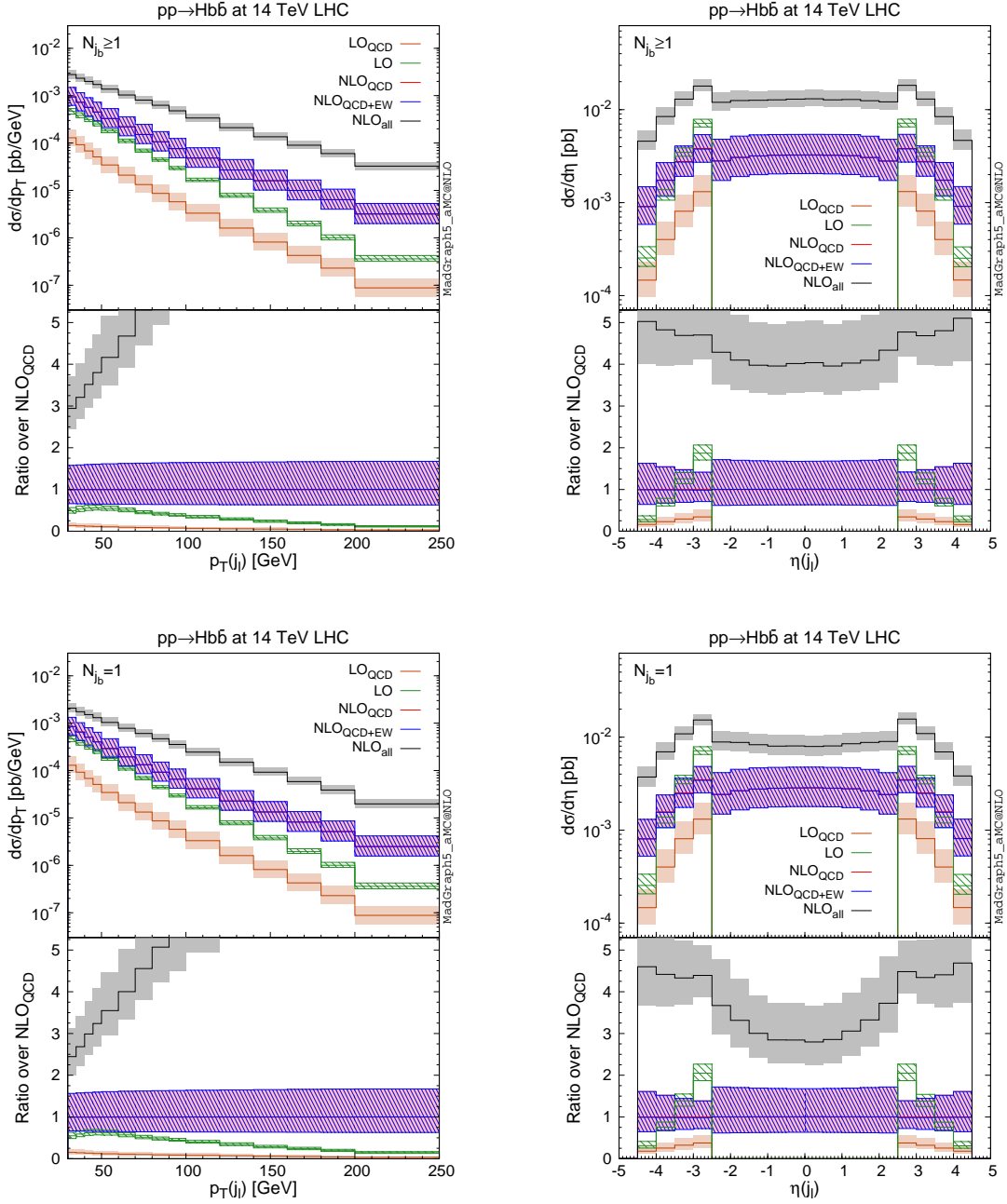

\centering
\includegraphics[page=\numexpr 6 \relax, width=.45\textwidth,draft=false]{pp_Hbb_GE1BJ_MSbar2.pdf}
\includegraphics[page=\numexpr 7 \relax, width=.45\textwidth,draft=false]{pp_Hbb_GE1BJ_MSbar2.pdf}\\
\vspace*{-10mm}
\includegraphics[page=\numexpr 6 \relax, width=.45\textwidth,draft=false]{pp_Hbb_EQ1BJ_MSbar2.pdf}
\includegraphics[page=\numexpr 7 \relax, width=.45\textwidth,draft=false]{pp_Hbb_EQ1BJ_MSbar2.pdf}\\
\vspace*{-10mm}
\caption{The $p_T(j_l)$ (left) and $\eta(j_l)$ (right) distribution for $N_{j_b}\ge1$ (up) and $N_{j_b}=1$ (down).}\label{fig:ptjl1}
\end{figure}

We now proceed to the analysis of differential distributions for several observables in the case $N_{j_b}\ge 1$ and $N_{j_b}=1$. First of all, beside documenting the result obtained, we want to explore the possibilities of enhancing the sensitivity on the $\sigma(y_b^2)$ contribution. In each of the Figs.~\ref{fig:ptjb1}--\ref{fig:ptjl1} we show distributions for a specific observable for $N_{j_b}\ge 1$ (upper plots) and $N_{j_b}= 1$ (lower plots) without (left plots) and with (right plots) the light-jet veto applied. We consider the following distributions: the transverse momenta and the pseudorapidity of the hardest $b$-jet, $p_T(j_{b,1})$ and $\eta(j_{b,1})$, the transverse momenta and the rapidity of the Higgs boson,  $p_T(H)$ and $y(H)$, the absolute value of the difference of the Higgs and hardest $b$-jet pseudorapidities, $|\Delta\eta(H,j_{b,1})|$,  and finally the transverse momenta and the pseudorapidity of the light-jet, $p_T(j_{l})$ and $\eta(j_{l})$. Since the last two observables are not defined in the case of the light-jet veto, we combine them in Fig.~\ref{fig:ptjl1}.

All the plots of Figs.~\ref{fig:ptjb1}--\ref{fig:ptjl1} have the same layout of those in Fig.~\ref{fig:mbb}, which has already been described. First of all, one can see that also at the differential level the $\NLO_2^\MSbar$ contribution, which is equal to the difference between the $\NLOQCDEW$ and $\NLOQCD$ predictions, is negligible. In absolute value, it reaches at maximum few percents of the $\NLOQCD$ prediction in the tails of the transverse-momentum distributions. For this reason, the $\CNLO/\NLOQCD$ ratio can be interpreted as the differential version of the ratio $[\sigma(y_b^2) +\sigma(\kappa_Z^2)]/\sigma(y_b^2)$, namely the inverse of the quantity displayed in the first column of Tab.~\ref{tab:couplings}. The higher is the $\CNLO/\NLOQCD$ ratio, the smaller is the fraction of the cross section that depends on $y_b$. The most important result that can be obtained by the analysis of all these plots is that whenever we look at phase-space regions that do {\it not} correspond to the bulk of the cross-section (large values of $p_T$, $|\eta|$ or $|y|$, {\it etc.}),  the $\CNLO/\NLOQCD$ ratio increases. In other words, applying cuts that depend on any of the observable we have considered, total rates diminish and at the same time the fraction of the cross section that depend on $y_b$ decreases. The only exception is the $|\Delta\eta(H,j_{b,1})|$ distribution, especially when the light-jet veto is applied. However, in order to halve the relative impact of $\sigma(\kappa_Z^2)$ term and bring it to roughly 30-40\% of $\sigma(y_b^2)$, rates have to be suppressed by a factor of 10. Thus, no real improvement can be gained. In conclusion, the sensitivity on $y_b$ cannot be improved even via the information at the differential level.

Although the main message of our phenomenological analysis has already been conveyed, we now report the other important features of plots in Figs.~\ref{fig:ptjb1}--\ref{fig:ptjl1}. We start with the $p_T(j_{b,1})$ distribution in Fig.~\ref{fig:ptjb1}. By comparing $\NLOQCD$ and $\LOQCD$ predictions one can see that the relative impact of the $\NLO_1^\MSbar|_{y_t=0}$ contribution is rather flat if the light-jet veto is not applied,  both in the $N_{j_b}\ge 1$ and $N_{j_b}=1$ cases. By applying the light-jet veto, the $\NLO_1^\MSbar|_{y_t=0}$ term becomes negative at large $p_T(j_{b,1})$ values, with a larger impact for the case $N_{j_b}\ge 1$. Both with and without the light-jet veto, $\NLOQCD$ scale uncertainties are much smaller than the $\LOQCD$ ones, also at the differential level.  As already said, the $\NLO_2^\MSbar$ contribution, which is equal to the difference between the $\NLOQCDEW$ and $\NLOQCD$ predictions, is negligible as in any other distribution. The $\LO$ prediction, which includes the $\LO_3$ term, is larger than the $\NLOQCD$  prediction, in particular the $\LO/\NLOQCD$ ratio grows for large  $p_T(j_{b,1})$ values and especially for $N_{j_b}\ge 1$ and/or applying the light-jet veto. It is important to note that in the case of the light-jet veto this effect is due to the suppression of the $\NLOQCD$ prediction; the $\LO$ prediction is only mildly affected by the light-jet veto also for large $p_T(j_{b,1})$. Moving to the $\CNLO$ prediction, which in particular includes the $\NLO_3$ term, also this quantity is larger than the $\NLOQCD$  prediction, and also the $\CNLO/\NLOQCD$ ratio grows for large  $p_T(j_{b,1})$ values, especially for $N_{j_b}\ge 1$. On the other hand, the impact of the light-jet veto is the opposite than in the $\LO$ case; the $\CNLO$ prediction is strongly reduced, especially at large  $p_T(j_{b,1})$ values. This is not surprising, since the VBF topology typically displays a light jet and therefore is completely suppressed. 

In the case of the $\eta(j_{b,1})$ distribution, Fig.~\ref{fig:etajb1}, similar considerations to the ones discussed for the $p_T(j_{b,1})$ distribution apply. The only difference is that the $\LOQCD/\NLOQCD$ ratio  is flat, also applying the light-jet veto, while the $\LO/\NLOQCD$ and $\CNLO/\NLOQCD$ ratios are mildly enhanced (suppressed) with (without) the light-jet veto in the peripheral region.

We now move to the Higgs boson distributions, starting with  $p_T(H)$ in Fig.~\ref{fig:ptH}. The peak of the distribution is at $p_T(H)\sim30$ GeV, since by definition $p_T(j_b)>30$ GeV. For the region  $p_T(H)>30$ GeV, the same considerations we have given for the $p_T(j_{b,1})$ distribution in Fig.~\ref{fig:ptjb1} apply also here. The situation is instead different for $p_T(H)\leq30$ GeV. Indeed, the $\LOQCD$ prediction is smaller than the $\NLOQCD$ one and strongly decreases close to the threshold, especially for the case $N_{j_b}= 1$. This is a pathological behaviour that is typical of fixed-order calculations in the presence of hard cuts.\footnote{In Ref.~\cite{Deutschmann:2018avk}, a larger bin width has been used in the distributions, hiding therefore the fixed-order  pathological behaviour for  $p_T(H)\leq30$ GeV.}  If at the same time $N_{j_b}= 1$,  $p_T(j_{b,1})>30$ GeV and $p_T(H)\leq30$ GeV, at $\LOQCD$ and more in general at $\LO$, the $b$-jet $j_{b,1}$ corresponds to a single $b_1$ quark/antiquark and the other $b_2$ antiquark/quark must have a momentum such that $\vec p_T(j_{b,1})+ \vec p_T(H)+ \vec p_T(b_2)=0$, where $\vec p_T$ denotes the azimuthal components of the momentum. Besides, the condition $p_T(b_2)$ < 30 GeV and/or $|\eta(b_2)|>2.5$ must be satisfied, otherwise $b_2$ would form another $b$-jet. These requirements all together pose  strong constraints on the $b_2$ phase-space, especially for $p_T(H)\TO 0$, suppressing the $\LOQCD$ and $\LO$ predictions. By adding a new particle in the final state, as in any NLO prediction, these hard cuts are removed and the pathological behaviour disappears. We also notice that the $\LO$ prediction, not the $\LOQCD$ one, considerably increases in this region moving from the $N_{j_b}= 1$ to $N_{j_b}\geq 1$ case. This is due to the presence of the $ZH$ topology in the $\LO_3$ term. By allowing  more than one $b$-jet, the $\LO_3$ and in turn $\LO$ predictions can easily satisfy the relation $\vec p_T(j_{b,1})+ \vec p_T(H)+ \vec p_T(b_2)=0$. Indeed, the $b$ quarks emerging from the $Z$ boson decay are back-to-back in the $Z$ boson rest-frame and not so rarely with $p_T(b)>30$ GeV. This leads to the presence of two $b$-jets,  $N_{j_b}= 2$, which does not suppress so much the $\LO_3$ contribution and in turn the $\LO$ contribution w.r.t.~the  $N_{j_b}= 1$ case, as can also be seen in Tab.~\ref{tab:rates}. Instead, in the case of $\LOQCD$, which is dominated by  ``genuine'' $\Hbb$ topology, bottom quarks are typically emitted collinearly to the beam-pipe axis. In principle, also for the $\LOQCD$ case, the conditions $p_T(H)<30$ GeV, $p_T(j_{b,1})>30$ GeV and $N_{j_b} \geq 1$ could be satisfied when  $N_{j_b} = 2$. In practice, at variance with the  $\LO$ case, at $\LOQCD$ this condition leads to large suppressions of the cross sections, as can also be seen in Tab.~\ref{tab:rates}. In the case of $\CNLO$ predictions, the $ZH$ topology is present in combination with additional QCD or QED real emissions, therefore the enhancement w.r.t.~$\LOQCD$ prediction is even higher than in the $\NLOQCD$ or $\LO$ case. On the other hand, we notice also that while the $\NLOQCD$ and the $\CNLO$ predictions are reduced by the light-jet veto, again the $\LO$ one is not.

In the case of the $y(H)$ distribution, Fig.~\ref{fig:yH}, the most important feature is the growth of the $\LO/\NLOQCD$ and $\CNLO/\NLOQCD$ ratios for large $|y(H)|$ values, especially if the light-jet veto is not applied. This can be understood by the fact that the Higgs boson recoils against the $b \bar b$ pair and possibly an additional real emission. Therefore, at the partonic level, {\it i.e.}, before the convolution with the PDFs, at $\LO$ or $\LOQCD$ accuracy and for large values of $y(H)$ we have  $|y(H)|\sim|\eta(H)|=|\eta(b \bar b)|\sim|y(b \bar b)|$. However, while the $b \bar b $ pair stems from the $Z$ boson decays in the $ZH$ topology, and therefore the entire $b \bar b $ pair tends to move in the same direction, in the case of a boosted $Z$ boson, in the ``genuine'' $\Hbb$ topology the bottom quarks tend to be emitted collinearly to the beam-pipe axis and back-to-back to each other. Therefore, in the $\LO$ predictions, and especially in the $\CNLO$ one which can get a further boost from the real emissions, the large $y(H)$ region is more populated than in the $\NLOQCD$ predictions. The  light-jet veto reduces this effect, clearly more for the $\CNLO$ case.   

Figure \ref{fig:DHjb1} shows the $|\Delta\eta(H,j_{b,1})|$ distribution, which as we have already said is the only one that displays a reduction of the $\CNLO/\NLOQCD$ (and also $\LO/\NLOQCD$) ratio by moving away from the bulk of the cross section, {\it i.e.}, going towards  large  $|\Delta\eta(H,j_{b,1})|$ values. For small $|\eta(H)|$ values, where the cross section is the largest, the probability of having the hardest $b$-jet with small $|\eta(j_{b,1})|$ values is higher for the $ZH$ topology (the $\LO_3$ and $\LO$ contributions) than in the ``genuine'' $\Hbb$ topology (the $\LOQCD$ contributions), since in the latter  bottom quarks tend to be emitted collinearly to the beam-pipe axis and back-to-back to each other. Also, at large $|\eta(H)|$ values, in the case of the $ZH$ topology in the $\LO_3$ the Higgs boson mostly recoils against the $b \bar b$ pair with the bottom quarks moving in the same direction, while in the ``genuine'' $\Hbb$ topology of  $\LOQCD$ the two bottoms tend mostly to have opposite directions, leading to one of them having (in the partonic rest frame) the pseudorapidity larger than the one of the Higgs boson in absolute value and with opposite sign. This dynamics is the origin of a flatter  $|\Delta\eta(H,j_{b,1})|$  distribution for $\LOQCD$ and $\NLOQCD$ predictions w.r.t.~the $\LO$ ones. The presence of real emissions and the VBF topology flattens the distribution moving from $\LO$ to $\CNLO$ accuracy. The flattening is even stronger moving from the $N_{j_b} \geq 1$  to the $N_{j_b}= 1$ case, which reduces the $ZH$ contribution.   

We finally discuss Fig.~\ref{fig:ptjl1}, which displays the $p_T(j_l)$ distribution on the left and the $\eta(j_l)$ one on the right. In the case of $p_T(j_l)$, we see how going to large  $p_T(j_l)$ values, the $\LO$ contribution decreases w.r.t~the $\NLOQCD$ one. We recall the at $\LO$ the light jets are only given by bottom quarks with pseudorapidity larger than 2.5 in absolute value, while in the $\NLOQCD$ predictions they can be genuine light-jets, with no $b$-quark inside them. Therefore, by requiring large pseudorapidities it is more difficult to achieve large transverse momenta. On the contrary, going to large  $p_T(j_l)$ values, the $\CNLO$ contribution increases w.r.t~$\NLOQCD$ one. Indeed, the light-jet in the VBF topology would not diverge in the limit  $p_T(j_l)\TO 0$, at variance with those from real QCD (or QED) emissions. For this reason the  $p_T(j_l)$ spectrum at $\CNLO$ is much flatter than the one at $\NLOQCD$ accuracy. 
Moving to the $\eta(j_l)$ distributions, the right plots clearly display the fixed-order pathological behaviour  for this observable in our calculation. Indeed, in the region $|\eta(j_l)|<2.5$, the $\LOQCD$ and $\LO$ predictions are exactly equal to zero. This is the reason why $\NLOQCD$ and $\CNLO$ scale uncertainties are smaller outside of this region; in the range $|\eta(j_l)|<2.5$ are in fact ``LO-type'' predictions. It is interesting to note how the peak of the distribution at   $\LO$ and $\CNLO$ accuracy is in the region  $|\eta(j_l)|\gtrsim 2.5$, so the pseudorapidity coverage of the $b$-jet tagging has a non-trivial impact on the numbers obtained in our work.

\section{Conclusions}\label{sec:conclusions}
The precise measurement of the Higgs boson couplings is one of the major goals of the LHC program. In particular, in the case of the Higgs-fermion Yukawa couplings, 
this translates in the need of measuring the relevant production mechanisms and/or decay modes of the Higgs boson. For what concerns the bottom quark, in principle
the extraction of $y_b$ from the measurement of  $\Hbb$  production would be subject to less theoretical assumptions than the corresponding $H\to b \bar b$ decay, whose branching ratio depends on all the other Higgs decay modes. However, the measurement of $\Hbb$  production  is plagued by various backgrounds,
with very large rates. 

In this paper, by computing the QCD and EW complete-NLO predictions for $\Hbb$ production in the 4FS, we have shown that the irreducible backgrounds involving 
the $HZZ$ interactions completely
submerge the ``genuine'' $y_b^2$-dependent $\Hbb$ signal. Among these backgrounds, one has both contributions where the accompanying bottom quarks originate from a resonant decay, $ZH$ production with $Z \to b\bar b$, but also contributions with a non-resonant spectrum, namely the $b$-associated VBF topology. Both these classes of irreducible backgrounds have very large cross sections when compared
to the ``genuine'' $\Hbb$ signal, and because of the
different underlying structures, it is extremely complicated to reduce their rate without \emph{de facto} killing also the signal. On this respect, in our study, we
have considered different set of cuts, both on the number of $b$-tagged jets and possibly vetoing extra light-jet radiation; in all cases the aforementioned backgrounds
are at least as large as the signal. 

Once the other irreducible backgrounds are also taken into account, namely those coming from the $gg$F+$b \bar b$ topology depending on $y_t$, their sum overwhelms the signal by about 
one order of magnitude. Thus, we find that it is tremendously difficult, if not impossible, to directly extract the bottom-quark Yukawa $y_b$ via the measurement of SM $\Hbb$ production at the 
LHC. Unless $y_b$  is significantly enlarged by new physics, a scenario which is strongly disfavoured by the $H\TO b \bar b$ experimental measurements, even for BSM scenarios the direct determination of $y_b$ via this process seems to be hopeless at the LHC. We have also investigated several differential distributions and we have found that moving away from the bulk of the cross section, not only the rates but also the fraction of them  that depends on $y_b$ decreases.

We reckon
that our study is performed at fixed-order, and neglects parton-shower, hadronisation and detector effects. Taking into account these effects it would be possible to perform a more realistic simulation. However, doing so, one should also consider on top of the $\Hbb$ irreducible backgrounds also those for the targeted Higgs decays and especially the reducible backgrounds. In general, this will further reduce
the signal-over-background ratio. Thus, we do not expect our conclusions to be altered, rather reinforced.
Beside the case of the measurement of ``genuine'' $\Hbb$ production, results presented in this paper can also be relevant for the background estimation 
for other Higgs processes, in particular $HH$ production
with one Higgs boson decaying to $b$ quarks (see {\it e.g.} Ref.~\cite{Homiller:2018dgu}). Similarly, the $\Hbb$ final state has to be taken into account when precise predictions for inclusive Higgs boson production are calculated. Therefore, the calculation presented in this paper can be in principle exploited also for this purpose, however, in this case one should  pay attention to not double-count $ZH$ and VBF contributions, which can be computed at higher accuracy via dedicated calculations. We leave studies in this direction for future work. Moreover, regardless of our phenomenological findings, the $\Hbb$ process remains a key process for the improvement and understanding of techniques for the computation of higher-order corrections in QCD, theoretical developments for the combination 4FS and 5FS computations at different perturbative orders and, as shown in this paper, for a better insight in the structure of the renormalisation condition in the EW sector. 

We want to stress that our aim is not discouraging  experimental analyses targeting  the signatures emerging from $\Hbb$ production. Similarly, we believe that such signatures and the corresponding contributions from ``genuine'' $\Hbb$ production should be taken into account as any other process in global analyses and fits for the determination of Higgs couplings and properties. However, our findings point to the fact that the measurement of ``genuine'' $\Hbb$ production is extremely challenging at the LHC and, given the current $y_b$ constraints from $H\TO b \bar b$ decays, its impact on global fit is expected to be negligible. Needless to say, all our phenomenological discussion concerns the LHC, {\it not} the possible future colliders, where higher energies and higher luminosities may change completely the picture. 

Finally, beside the phenomenological results, we have extended the capabilities of the \mgshort{} framework in order to have the
possibility to compute NLO EW corrections and more in general QCD and EW complete-NLO predictions in the 4FS. This feature will be included in a future release of the code. To the best of our knowledge, this 
work is the first in which NLO EW corrections or complete-NLO predictions are computed in the 4FS for a complete process 
at hadron colliders. In the case of $\Hbb$ production, this work represents the first ever full computation of NLO EW corrections and complete-NLO predictions. While the impact of NLO EW corrections is found to be very small, once again complete-NLO predictions turns out to be much larger than naively expected values, due to the presence of new topologies, in this case the $ZH$ and VBF ones.

\section*{Acknowledgements}

    We want to thank for interesting discussions and suggestions Sasha Nikitenko, Frank Tackmann, Maria Ubiali and Marius Wiesemann. We are grateful to
    the developers of {\sc MadGraph5\_aMC@NLO} for the long-standing collaboration and for discussions.
    
   The work of D.~P.~is supported by the Deutsche Forschungsgemeinschaft (DFG) under Germany's Excellence Strategy - EXC 2121 ``Quantum Universe'' - 390833306. H.S.S is supported by the ILP Labex (ANR-11-IDEX-0004-02, ANR-10-LABX-63).

\bibliographystyle{utphys}
\bibliography{hbbew}

\end{document}